**An improved Epidemiological-Unscented Kalman Filter (Hybrid SEIHCRDV-UKF) model for the prediction of COVID-19. Application on real-time data.**

Vasileios E. Papageorgiou[1], George Tsaklidis[2]
Department of Mathematics, Aristotle University of Thessaloniki, Thessaloniki 54124, Greece
Correspondence: tsaklidi@math.auth.gr[2]; vpapageor@math.auth.gr[1]

**Abstract**

The prevalence of COVID-19 has been the most serious health challenge of the 21th century to date, concerning national health systems on a daily basis, since December 2019 when it appeared in Wuhan City. Nevertheless, most of the proposed mathematical methodologies aiming to describe the dynamics of an epidemic, rely on deterministic models that are not able to reflect the true nature of its spread. In this paper, we propose a SEIHCRDV model – an extension/improvement of the classic SIR compartmental model – which also takes into consideration the populations of exposed, hospitalized, admitted in intensive care units (ICU), deceased and vaccinated cases, in combination with an unscented Kalman filter (UKF), providing a dynamic estimation of the time dependent system's parameters. The stochastic approach is considered necessary, as both observations and system equations are characterized by uncertainties. Apparently, this new consideration is useful for examining various pandemics more effectively. The reliability of the model is examined on the daily recordings of COVID-19 in France, over a long period of 265 days. Two major waves of infection are observed, starting in January 2021, which signified the start of vaccinations in Europe providing quite encouraging predictive performance, based on the produced NRMSE values. Special emphasis is placed on proving the non-negativity of SEIHCRDV model, achieving a representative basic reproductive number $R_0$ and demonstrating the existence and stability of disease equilibria according to the formula produced to estimate $R_0$. The model outperforms in predictive ability not only deterministic approaches but also state-of-the-art stochastic models that employ Kalman filters. Furthermore, the relevant analysis supports the importance of vaccination, as even a small increase in the dialy vaccination rate could lead to a notable reduction in mortality and hospitalizations.

## 1. Introduction

The novel coronavirus SARS-CoV-2 (COVID-19) had broken out in the city of Wuhan, China during December 2019 (Muralidar et al. 2020). The pathogenesis of virus is characterized by respiratory tract infection, which can lead to pneumonia showing ground glass alveolar angiography. This highly contagious disease has been declared a pandemic from the World Health Organization since January 2020, while a year later, the virus has infected more than 100 million people (Coronavirus Research Center of John Hopkins University 2021), despite various health measures taken by many national governments.

The severity of the situation along with the need for early prevention, led to an investigation into the nature of COVID through mathematical modelling. Usually, the spread of contagious diseases is described using compartmental models, where SIR (susceptible-infected-recovered) represents the most widely known model (Brauer et al. 2019). As a result, many articles base their exploration of COVID dynamics on the SIR model (Cooper I. et al 2020), or some of its extensions such as SIRS (susceptible-infected-recovered-susceptible) (Salman A.M. et al 2021), the SEIR (susceptible-exposed-infectious-recovered) (He et al. 2020) or the SEIRD (susceptible-exposed-infectious-recovered-deceased) models (Rajagopal K. et al 2020). Moreover, Malkov 2020 suggested a SEIRS (susceptible-exposed-infectious-resistant-susceptible) model with time-varying transmission rates.

However, these approaches are not able to describe the time-varying dynamics of the pandemic, especially over long time periods, due to various fluctuations in the disease parameters. For example, the establishment of restrictive measures such as the use of masks in closed and public environments or the onset of lockdown in many countries around the world, reduced daily infection rates (Atalan 2020), while the emergence of variants such as B.1.1.7 (alpha) or B.1.617.2 (delta) are associated with increased contagiousness (Kidd et al. 2021), increased risk of hospitalization and ICU admission, as well as increased mortality (Tuite et al. 2021; Veneti et al. 2021; Davies et al. 2021). In addition, Twohig et al. (2021), state that there is a higher risk of serious illness for the unvaccinated individuals infected with delta than the alpha variant.

Since the asymptomatic infections are difficult to detect, and in addition to the false positive-negative ratio of the PCR test, we realize the existence of potential uncertainties in the reported data (Hu et al. 2020, Keeling et al. 2020). Furthermore, the states and corresponding transitions of the majority of compartmental models do not take into account the complete dynamics of the disease. Therefore, the transition from a deterministic to a stochastic approach seems necessary. Singh et al. (2021) utilize the standard Kalman Filter to estimate the evolution of the pandemic in India, but as they state, these estimations are only reliable over a short time period. Costa et al. (2005) combine a SEIR model with an extended Kalman filter (EKF) to simulate the outbreak of an epidemic, while Ndanguza et al. (2016) include in the SEIR-EKF combination the estimation of the epidemiological model's parameters. Sebbagh and Kechida (2022) propose a SIRD-EKF model for the estimation of the evolution of COVID-19. Zhu et al. (2021) propose an SEIRD-EKF model with dynamic parameter estimation. Song et al. (2021) combine the same two models, while the parameter estimation of the SEIRD is based on an iterative optimization method based on maximum likelihood. Other attempts, deploy the ensemble Kalman Filter in combination with some extensions of the SIR model (Nkwayep et al. 2022; Lal et al. 2021), while Calvetti et al. (2021) adopt a SEAIR (susceptible-exposed-asymptomatic-infectious-recovered) model and utilize the particle filtering methodology to estimate the spread of the virus in Ohio and Michigan. Finally, Marioli et al. (2021) estimate the evolution of $R_0$ using Kalman filtering.

In this paper, we propose a novel hybrid SEIHCRDV (susceptible-exposed-infectious-hospitalized-ICU admitted-recovered-deceased-vaccinated) model with an unscented Kalman filter (UKF) with dynamic parameter estimation. This approach can effectively explain significant fluctuations in the spread of COVID after the start of the vaccination season, providing a much more representative picture of the daily evolution of the pandemic, due to the –appropriately-increased number of suitable states and associated transitions. The choice of the UKF enriches the existing relevant literature, where only the EKF is applied, while we highlight the advantages of the estimations produced by this choice.

The increase in the number of mixed differential equations, where the majority of states (6 out of 8) are observable, combined with the dynamic parameter estimation resulting from the real time feedback of observations, prevents the occurrence of extreme parameter estimations, giving quite reliable predictions even for the hidden states of the model. The proposed methodology can be easily implemented, in cases where both the observations and the system suffer from uncertainties. We aim to eliminate unnecessary noise, providing robust estimations for both observable and hidden states. The inclusion of the hospital and ICU admitted states in the model, provides another important advantage to the analysis. COVID-19 is characterized by a high percentage of asymptomatic cases, leading to the consideration that the levels of daily infective and recovered cases could be deemed as indices of low trustworthiness for the evolution of the epidemic. On the contrary, hospital and ICU admitted cases are tested/recorded thoroughly for confirmation of COVID infection, rendering the daily observations of these two states as the most accurate indicators for assessing the fitting-predictive capacity of the proposed model. It should be noted that the addition of the aforementioned states does not impose significant computational burden to the model. This also helps to reduce the possibility of overfitting that may occur in any statistical or artificial intelligence model (Papageorgiou 2021).

In addition, a mathematical analysis is carried out emphasizing the non-negativity of the system states, as long as the existence and stability of disease-free and endemic equilibria. We even propose an alternative-improved formula for index $R_0$ – usually referred to as the basic reproduction number – based on the proposed compartmental model, drawing valuable conclusions about the nature and the future of the disease. We test the reliability of our model on reported data from France, covering a long period of 265 days, providing daily estimates for the infectious, hospitalized, ICU admitted, recovered, deceased and vaccinated cases. Finally, the additional states are carefully selected aiming to include only states with available daily observations. Specifically, apart from the susceptible and exposed states, the remaining 6 states that are part of the proposed model are accompanied by observations available on a daily basis. We efficiently employ this supplementary information in our model, greatly enhancing its fitting and predictive capacity. Thus, unlike the majority of the existing literature, we not only propose a novel stochastic epidemiological model and examine its predictive ability, but also explore its mathematical properties like the existence and stability of epidemic equilibria.

The article is organized as follows: In section 2, we present the mathematical structure of the hybrid SEIHCRDV-UKF model in parallel with the theoretical results based on this compartmental model. In section 3, we provide simulations for varying model parameters, and test its fitting-predictive performance to the daily observations of France, starting in January of 2021 where we encounter the first reported fully vaccinated cases. In section 4 we discuss and in section 5 we conclude the main findings of our analysis with particular emphasis on the advantages derived from our stochastic model.

## 2. Mathematical tools and Methods

### *2.1. Proposed SEIHCRDV epidemiological model*

In this paragraph, we present the SEIHCRDV compartmental model, through which we then examine the spread of coronavirus in France using real data. As previously mentioned, we propose an extension of the standard SIR model, introducing five additional states and associated differential equations to account for the exposed, hospitalized, ICU admitted, deceased and vaccinated cases. Figure 1 displays the transitions between the states of the proposed epidemiological model. The selection of presented states and transitions has been considered with great care, aiming to manage the complexity of the examined phenomenon without significantly increasing the complexity of the model.

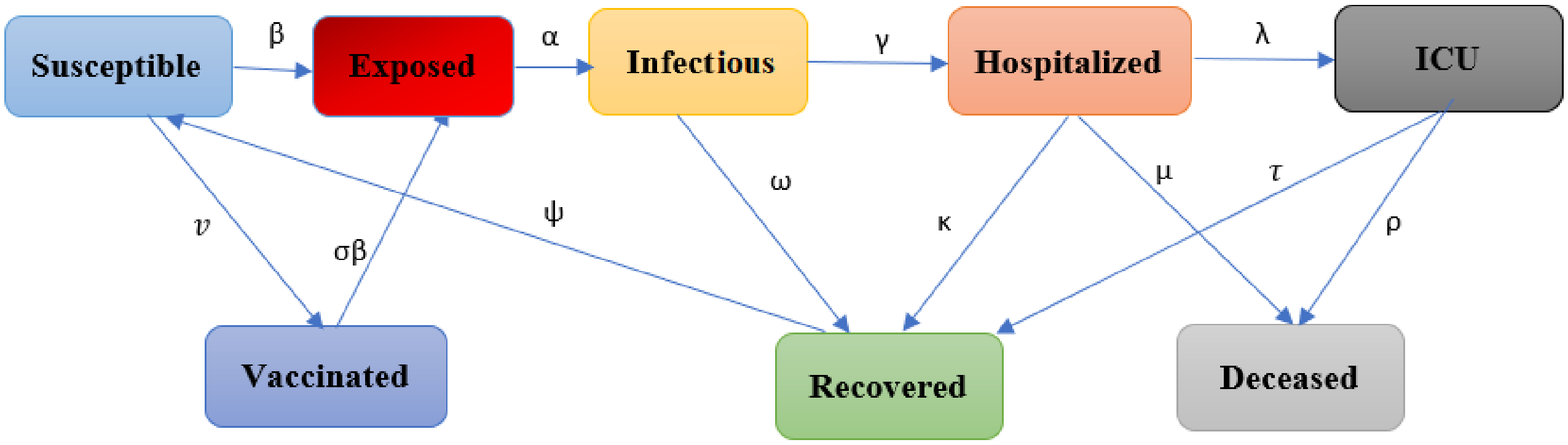

**Figure 1.** Diagram of the SEIHCRDV epidemiological model

Notice that system (2) provides a novel deterministic model of the form

$$\dot{\boldsymbol{X}}(t) = f(t, \boldsymbol{X}) \quad (1)$$

that describes the evolution of COVID-19 spread in the population. We denote with $S(t), E(t), I(t), H(t), C(t), R(t), D(t)$ and $V(t)$ the susceptible, exposed, infectious, hospitalized, placed in intense care units (ICU), recovered, deceased and vaccinated cases, respectively, that change over time $t$. The notation $\beta$ denotes the infection rate, $\alpha$ the incubation rate, $\gamma$ the hospitalization rate, $\lambda$ the admission rate in ICU, $\kappa$ and $\tau$ the recovery rates of hospitalized and ICU admitted patients respectively, $\mu$ and $\rho$ the mortality rates of hospitalized and ICU admitted patients respectively, $v$ the vaccination rate, $\sigma$ the vaccine breakthrough infection rate of fully vaccinated individuals, $\psi$ the re-susceptibility rate and $\omega$ the transition rate from the infectious to the recovered state that can be considered constant over time (Table 1). Moreover, $\gamma$, $\alpha$ and $\sigma$ can be also deemed as constants, while $1-\sigma$ displays the average protection provided by complete vaccination.

System (2) that follows, presents analytically the structure and evolution of our model,

$$\frac{dS(t)}{dt} = -\frac{\beta S(t) I(t)}{N} - vS(t) + \psi R(t)$$

$$\frac{dE(t)}{dt} = \frac{\beta S(t) I(t)}{N} + \frac{\sigma \beta V(t) I(t)}{N} - aE(t)$$

$$\frac{dI(t)}{dt} = aE(t) - \gamma I(t) - \omega I(t)$$

$$\frac{dH(t)}{dt} = \gamma I(t) - \lambda H(t) - \kappa H(t) - \mu H(t)$$

$$\frac{dC(t)}{dt} = \lambda H(t) - \tau C(t) - \rho C(t) \tag{2}$$

$$\frac{dR(t)}{dt} = \kappa H(t) + \tau C(t) + \omega I(t) - \psi R(t)$$

$$\frac{dD(t)}{dt} = \mu H(t) + \rho C(t)$$

$$\frac{dV(t)}{dt} = vS(t) - \frac{\sigma\beta V(t) I(t)}{N}$$

where we consider population size $N = S(t) + E(t) + I(t) + H(t) + C(t) + R(t) + D(t) + V(t)$ to be constant during the pandemic. Consequently, we get that

$$\frac{dN}{dt} = 0 = \frac{dS(t)}{dt} + \frac{dE(t)}{dt} + \frac{dI(t)}{dt} + \frac{dH(t)}{dt} + \frac{dC(t)}{dt} + \frac{dR(t)}{dt} + \frac{dD(t)}{dt} + \frac{dV(t)}{dt},$$

a condition that is satisfied by the eight above ordinary differential equations (ODEs) that describe the evolution of the pandemic inside the population. An extension of model (2) can be proposed by assuming that the population $N(t)$ is open, adding to model (2) a representative birth-migration rate $\Lambda$ and an extra mortality rate $\delta$, due to natural, non-COVID causes.

Both $\Lambda$ and $\delta$ should be considered constant for relatively short time periods. The consideration of these two extra parameters leads to the following model,

$$\frac{dS(t)}{dt} = \Lambda - \frac{\beta S(t) I(t)}{N} - vS(t) + \psi R(t) - \delta S(t)$$

$$\frac{dE(t)}{dt} = \frac{\beta S(t) I(t)}{N} + \frac{\sigma\beta V(t) I(t)}{N} - aE(t) - \delta E(t)$$

$$\frac{dI(t)}{dt} = aE(t) - \gamma I(t) - \omega I(t) - \delta I(t)$$

$$\frac{dH(t)}{dt} = \gamma I(t) - \lambda H(t) - \kappa H(t) - \mu H(t) - \delta H(t)$$

$$\frac{dC(t)}{dt} = \lambda H(t) - \tau C(t) - \rho C(t) - \delta C(t) \tag{3}$$

$$\frac{dR(t)}{dt} = \omega I(t) + \kappa H(t) + \tau C(t) - \psi R(t) - \delta R(t)$$

$$\frac{dD(t)}{dt} = \mu H(t) + \rho C(t)$$

$$\frac{dV(t)}{dt} = vS(t) - \frac{\sigma\beta V(t) I(t)}{N} - \delta V(t)$$

which is utilized throughout the rest of our study.

Apparently, the features of the continuous time formulation have to be derived using the discrete time observations of real-time data, rendering the discretization of the system a necessary step of this analysis. System (3) can be discretized using finite differences, namely $\frac{df(t)}{dt} = \frac{f_{k+\Delta t} - f_k}{\Delta t}$, producing system (4) that is a transformation of the ODEs system of 8 equations into a state-space representation. As a result, the following system

$$S_{k+\Delta t} = \Lambda\Delta t + \left(1 - \Delta t\left(\frac{\beta I_k}{N} + \delta + v\right)\right) S_k + \psi R_k \Delta t$$

$$E_{k+\Delta t} = (1 - (a + \delta)\,\Delta t)E_k + \frac{\beta S_k I_k}{N}\Delta t + \frac{\sigma\beta I_k V_k}{N}\,\Delta t$$
$$I_{k+\Delta t} = (1 - \Delta t(\gamma + \omega + \delta))\,I_k + \Delta t\,\alpha E_k$$
$$H_{k+\Delta t} = \big(1 - \Delta t(\lambda + \kappa + \mu + \delta)\big)H_k + \Delta t\,\gamma I_k$$
$$C_{k+\Delta t} = (1 - \Delta t(\tau + \rho + \delta))C_k\ + \Delta t\,\lambda H_k \tag{4}$$
$$R_{k+\Delta t} = (1 - \Delta t(\psi + \delta))R_k + \Delta t\,\omega I_k + \Delta t\,\kappa H_k + \Delta t\,\tau C_k$$
$$D_{k+\Delta t} = D_k + \Delta t\,\mu H_k + \Delta t\,\rho C_k$$
$$V_{k+\Delta t} = \left(1 - \Delta t(\frac{\sigma\beta}{N} I_k + \delta)\right)V_k + \Delta t\,\nu S_k$$

**Table 1.** Parameter and state definition of the proposed SEIHCRDV model

| Symbol | Definition of Parameter/State |
|---|---|
| S | Susceptible |
| E | Exposed |
| I | Infectious |
| H | Hospitalized |
| C | ICU admitted |
| R | Recovered |
| D | Deceased |
| V | Vaccinated |
| Λ | Birth and migration rate |
| $\alpha$ | Incubation rate |
| $\beta$ | Infection rate |
| $\gamma$ | Rate of hospitalizations |
| $\delta$ | Mortality rate due to non-COVID causes |
| $\lambda$ | Transition rate from hospital to ICU |
| $\kappa$ | Recovery rate of hospitalized cases |
| $\mu$ | Death rate of hospitalized cases |
| $\nu$ | Fully vaccination rate |
| $\rho$ | Death rate of ICU admitted cases |
| $\sigma$ | Vaccine breakthrough infection rate of fully vaccinated cases |
| $\tau$ | Recovery rate of ICU admitted cases |
| $\psi$ | Re-susceptibility rate |
| $\omega$ | Recovery rate of infected cases with mild symptoms |

is more suitable for interpreting the evolution of COVID during discrete time steps, while the states in (4) at the $k$-th time point can be assembled in a vector $\boldsymbol{x}_k^T = [S_k\ E_k\ I_k\ H_k\ C_k\ D_k\ V_k]$. Some constant parameters that accompany the above discretized system could be deemed as time-varying, resulting in an augmented state vector $\boldsymbol{x}_k^T = [S_k\ E_k\ I_k\ H_k\ C_k\ D_k\ V_k\ \boldsymbol{\theta}_k^T]$, where $\boldsymbol{\theta}_k^T$ is a vector containing the parameters that can be updated over time.

We notice that in both models (2) and (3), there are no transitions directly linking infectious and deceased cases. Moreover, these models contain a differential equation for the vaccinated cases, providing a representation that is more appropriate in the later stages of the pandemic, namely from January of 2021 onwards. The period between the onset of COVID-19 in Wuhan and the start of the vaccination period, was sufficient to understand the risks of the disease and to inform citizens about the seriousness of the situation. Hence, we believe that the proportion of severely infected people who died of COVID-19 without being admitted to a hospital, or an ICU is negligible. In addition, the negligible effect of this transition could be deemed as part of the Gaussian noise added to the state equations, resulting in the stochastic equivalents of the above SEIHCRDV models (2) and (3).

We should pay attention to the value of the re-susceptibility rate $\psi$. It is known that in COVID, the immunity provided by an infection is temporary. However, many epidemiological models do not consider a transition from the recovered to the susceptible state. Some previous studies report that there is no significant evidence regarding the ideal selection of this rate (Salman et al. 2021; Malkov 2020). Although, based on research in infected people with mild symptoms, the proportion of virus-neutralizing antibodies is stable for up to 6 months, providing immunity to re-infection.

*2.2 Mathematical analysis*

**Theorem 1.** Given that the initial state sizes of the SEIHCRDV system (3) are non-negative, i.e. $S_0, E_0, I_0, H_0, C_0, R_0, D_0, V_0 \geq 0$, then the respective trajectories remain non-negative for all $t > 0$.

**Proof.** Firstly, we note that all 14 transition rates of the SEIHCRDV system (3) are non-negative constants. From the second equation of the SEIHCRDV system, we get that

$$E'(t) = \frac{\beta S(t) I(t)}{N} + \frac{\sigma \beta V(t) I(t)}{N} - aE(t) - \delta E(t) \geq -(a+\delta)E(t) \tag{5}$$

or

$$(\ln E(t))' \geq -(a+\delta). \tag{6}$$

By integrating inequality (6), we derive

$$E(t) \geq e^{-(a+\delta)\int dt} = c_1 e^{-(a+\delta)t} \tag{7}$$

and thus

$$E(t) \geq E_0 e^{-(a+\delta)t} \geq 0. \tag{8}$$

Following similar methodology, we can prove the non-negativity of $I(t), H(t), C(t), R(t)$ and $D(t)$. Next, we prove the non-negativity of the function of vaccinated cases $V(t)$, using the infinity norm $\left||f|\right|_\infty = \max_{t \in D_f} |f(t)|$ thus,

$$V'(t) = vS(t) - \frac{\sigma \beta V(t) I(t)}{N} - \delta V(t) \geq -\left(\frac{\sigma \beta I(t)}{N} + \delta\right) V(t) \geq -\left(\frac{\sigma \beta \max_{t \in [0,\infty)} |I(t)|}{N} + \delta\right) V(t)$$

and consequently,

$$\ln\big(V(t)\big)' \geq -\frac{\sigma \beta \left||I|\right|_\infty}{N} - \delta.$$

By integrating the above inequality and substituting for $t = 0$, we culminate in

$$V(t) \geq V_0 e^{-\left(\frac{\sigma \beta \left||I|\right|_\infty}{N} + \delta\right)t} \geq 0. \tag{9}$$

In a similar manner, we can show the non-negativity of susceptible cases. Notice that $\left||I|\right|_\infty < \infty$, as we refer to a finite population function $N(t)$. ■

**Theorem 2.** The basic reproduction number of the proposed SEIHCRDV model (3) is

$$R_0 = \frac{\beta \alpha (S^0 + \sigma V^0)}{N(\alpha+\delta)(\gamma+\omega+\delta)} = \frac{\beta \alpha \Lambda (\delta + \sigma v)}{N\delta(\alpha+\delta)(\gamma+\omega+\delta)(v+\delta)}.$$

**Proof.** We base our analysis on the utilization of the next-generation matrix proposed by van den Driessche and Watmough (2002) for the definition of the basic reproduction number $R_0$. Coefficient $R_0$ is defined as the number of new infected cases produced by another already infected individual belonging to the population.

Let $\boldsymbol{X} = (E, I, H, C)^T$ be the vector consisting of the 4 states that represent infected cases, and

$$\boldsymbol{Y}^0 = (S^0, 0,0,0,0,0, V^0) = \left(\frac{\Lambda}{v+\delta}, 0,0,0,0,0, \frac{v\Lambda}{\delta(v+\delta)}\right), \tag{10}$$

be the disease-free equilibrium (DFE), where all individuals are gathered in the susceptible and vaccinated states. This DFE is obtained after equating the ODEs of system (3) with zero, namely $f(t, X) = 0$, while setting $I = 0$ (Brauer et al. 2019). We set $\boldsymbol{X}^0 = \boldsymbol{0}^T$, as there should be no cases in any of the infected states of our model during the disease-free period.

Let $\mathcal{F}(\boldsymbol{X})$ and $\mathrm{V}(\boldsymbol{X})$ be the 4x1 vectors whose $i$-th entry exhibits the rate of new infections entering state $i$ and the transfer rate out of state $i$ by non-infection means, respectively. Moreover, we do not consider the transition of an individual between the 4 states as a new infection, but rather the progression of the infected individual through the various states of the system. As a result,

$$\dot{\boldsymbol{X}}^T = \mathcal{F}(\boldsymbol{X}) - \mathrm{V}(\boldsymbol{X}), \tag{11}$$

where

$$\mathcal{F}(\boldsymbol{X}) = \left(\frac{\beta SI}{N} + \frac{\sigma\beta VI}{N}, 0,0,0\right)^T$$

and

$$\mathrm{V}(\boldsymbol{X}) = ((a+\delta)E, (\gamma+\omega+\delta)I - aE, (\lambda+\kappa+\mu+\delta)H - \gamma I, (\tau+\rho)C - \lambda H)^T.$$

The next-generation matrix is defined as the matrix product $\boldsymbol{F}\boldsymbol{V}^{-1}$ where matrices $\boldsymbol{F}, \boldsymbol{V}$ are the Jacobians of $\mathcal{F}(\boldsymbol{X}), \mathrm{V}(\boldsymbol{X})$ evaluated at the disease-free equilibrium $\boldsymbol{Y}^0$ (10). Hence, both $\boldsymbol{F}$ and $\boldsymbol{V}$ are 4x4 dimensional matrices of the forms

$$\boldsymbol{F}_{|Y^0} = \begin{pmatrix} 0 & \frac{\beta S^0 + \sigma\beta V^0}{N} & 0 & 0 \\ 0 & 0 & 0 & 0 \\ 0 & 0 & 0 & 0 \\ 0 & 0 & 0 & 0 \end{pmatrix}, \tag{12}$$

and

$$\boldsymbol{V}_{|Y^0} = \begin{pmatrix} \delta + a & 0 & 0 & 0 \\ -a & (\gamma+\omega+\delta) & 0 & 0 \\ 0 & -\gamma & (\lambda+\kappa+\mu+\delta) & 0 \\ 0 & 0 & -\lambda & (\tau+\rho+\delta) \end{pmatrix}, \tag{13}$$

respectively, while $\boldsymbol{V}$ is an invertible lower triangular matrix, with a non-zero determinant equal to the product of its diagonal elements. Then, we get that

$$\boldsymbol{V}^{-1} = \begin{pmatrix} \frac{1}{\alpha+\delta} & 0 & 0 & 0 \\ \frac{\alpha}{(\alpha+\delta)(\gamma+\omega+\delta)} & \frac{1}{\gamma+\omega+\delta} & 0 & 0 \\ \frac{\alpha\gamma}{(\alpha+\delta)(\kappa+\lambda+\mu+\delta)(\gamma+\omega+\delta)} & \frac{\gamma}{(\kappa+\lambda+\mu+\delta)(\gamma+\omega+\delta)} & \frac{1}{\kappa+\lambda+\mu+\delta} & 0 \\ \frac{\alpha\gamma\lambda}{(\alpha+\delta)(\kappa+\lambda+\mu+\delta)(\gamma+\omega+\delta)(\tau+\rho+\delta)} & \frac{\gamma\lambda}{(\kappa+\lambda+\mu+\delta)(\gamma+\omega+\delta)(\tau+\rho+\delta)} & \frac{\lambda}{(\kappa+\lambda+\mu+\delta)(\tau+\rho+\delta)} & \frac{1}{(\tau+\rho+\delta)} \end{pmatrix}$$

while $R_0$ is the spectral radius of the next-generation matrix $\boldsymbol{F}\boldsymbol{V}^{-1}$, given by

$$R_0 = \rho(\boldsymbol{F}\boldsymbol{V}^{-1}) = \frac{\beta\alpha(S^0 + \sigma V^0)}{N(\alpha + \delta)(\gamma + \omega + \delta)} \quad (14)$$

where $\rho(\boldsymbol{A})$ stands for the spectral radius of matrix $\boldsymbol{A}$.

By substituting the DFE values in equation (14) we get that

$$R_0 = \frac{\alpha\beta\Lambda(\delta + \sigma\nu)}{N\delta(\alpha + \delta)(\gamma + \omega + \delta)(\nu + \delta)} \,. \quad (15)$$

Similarly, we can calculate the basic reproductive number for system (2) where we remove the impact of exogenous factors like new births and mortality due to causes not related to COVID, namely,

$$R_0 = \frac{\beta(S^0 + \sigma V^0)}{N(\gamma + \omega)} \,. \quad (16)$$

A dynamic equivalent of the basic reproduction number $R_0$ that can be calculated in a daily basis during the pandemic is the effective reproduction number $R_t$ that is represented by the formula

$$R_t = \frac{\beta(S_t + \sigma V_t)}{N(\gamma + \omega)}. \quad (17)$$

■

**Theorem 3.** The ratio of the evolution rates of deceased cases and ICU admissions is lower bounded, i.e., $\frac{dD(t)}{dt} / \frac{dC(t)}{dt} \geq \frac{\mu_t}{\lambda_t}$, during periods of positive ICU admissions rate.

**Proof.** Firstly, using the fifth and seventh equation of system (3), corresponding to the evolution of ICU admissions and deceased individuals through time, we have

$$\frac{dC(t)}{dt} \leq \lambda H(t), \quad \frac{dD(t)}{dt} \geq \mu H(t) \,. \quad (18)$$

Consequently,

$$\frac{dD(t)}{dt} \Big/ \frac{dC(t)}{dt} \geq \frac{\mu}{\lambda}. \quad (19)$$

while by considering the potential dynamic evolution of parameters $\mu$ and $\lambda$ we get that

$$\frac{dD(t)}{dt} \Big/ \frac{dC(t)}{dt} \geq \frac{\mu_t}{\lambda_t}. \quad (20)$$

■

**Theorem 4.** If $R_0 > 1$, then the proposed SEIHCRDV model leads asymptotically to a unique, non-trivial endemic equilibrium.

**Proof.** Let $\boldsymbol{Y}^* = (S^*, E^*, I^*, H^*, C^*, R^*, D^*, V^*)^T$ be the SEIHCRDV's endemic equilibrium that can be determined after setting all differential equations of system (3) equal to 0. We may limit the analysis to 7 equations, as $D(t)$ can be described as a linear combination of the remaining 7 states and does not contribute to any other differential equation.

We explore the existence and uniqueness of a non-trivial endemic equilibrium, as $\boldsymbol{Y^0}$ obviously constitutes the disease-free equilibrium for system (3).

For simplicity let $\beta_1 = \frac{\beta}{N} > 0$. Then,

$$
\begin{gathered}
\Lambda - \beta_1 S^* I^* - v S^* + \psi R^* - \delta S^* = 0 \\
\beta_1 S^* I^* + \sigma \beta_1 V^* I^* - a E^* - \delta E^* = 0 \\
\alpha E^* - \gamma I^* - \omega I^* - \delta I^* = 0 \\
\gamma I^* - \lambda H^* - \kappa H^* - \mu H^* - \delta H^* = 0 \\
\lambda H^* - \tau C^* - \rho C^* - \delta C^* = 0 \\
\omega I^* + \kappa H^* + \tau C^* - \psi R^* - \delta R^* = 0 \\
v S^* - \sigma \beta_1 V^* I^* - \delta V^* = 0 \,.
\end{gathered}
$$

According to the above non-linear system, we aim to describe each state in the equilibrium as a function of $I^*$, resulting in the formulas,

$$E^* = \frac{\gamma + \omega + \delta}{\alpha} I^* \tag{21}$$

$$H^* = \frac{\gamma}{\lambda + \kappa + \mu + \delta} I^* \tag{22}$$

$$C^* = \frac{\lambda \gamma}{(\lambda + \kappa + \mu + \delta)(\tau + \rho + \delta)} I^* \tag{23}$$

$$R^* = \frac{\omega(\lambda + \kappa + \mu + \delta)(\tau + \rho + \delta) + \kappa\gamma(\tau + \rho + \delta) + \tau\lambda\gamma}{(\psi + \delta)(\lambda + \kappa + \mu + \delta)(\tau + \rho + \delta)} I^* = c_1 I^* \tag{24}$$

$$S^* = \frac{\Lambda + \psi c_1 I^*}{\beta_1 I^* + v + \delta} \tag{25}$$

$$V^* = \frac{v\Lambda + v\psi c_1 I^*}{(\sigma\beta_1 I^* + \delta)(\beta_1 I^* + v + \delta)}. \tag{26}$$

It should be noticed here that $c_1 \in \left(0, \frac{\gamma+\omega}{\psi+\delta}\right)$. This statement can be easily proved by rewriting $c_1$ appearing in (24), in the form

$$c_1 = \frac{1}{\psi + \delta}\left[\omega + \frac{\gamma\,(\lambda\tau + \kappa(\tau + \rho + \delta))}{(\lambda + \kappa + \mu + \delta)(\tau + \rho + \delta)}\right] < \frac{\gamma + \omega}{\psi + \delta}\ , \tag{27}$$

as $\lambda\tau + \kappa(\tau + \rho + \delta) < (\lambda + \kappa + \mu + \delta)(\tau + \rho + \delta)$.

By substituting equations (21), (25) and (26) into the second equation of the system, we derive the cubic equation without constant coefficient,

$$
\begin{aligned}
a\beta_1(\Lambda + \psi c_1 I^*)(\sigma\beta_1 I^* + \delta) I^* + \sigma\beta_1 \alpha v(\Lambda + \psi c_1 I^*) I^* \\
-(a + \delta)(\gamma + \omega + \delta)(\sigma\beta_1 I^* + \delta)(\beta_1 I^* + \delta + v) I^* = 0\,,
\end{aligned} \tag{28}
$$

which has one zero and two non-zero solutions. The corresponding quadratic equation resulting from (28) to determine the two aforementioned solutions, is

$$
\begin{aligned}
&\sigma\beta_1^2[\alpha\psi c_1 - (a + \delta)(\gamma + \omega + \delta)] I^{*2} \\
&\quad + [\alpha\beta_1(\Lambda\beta_1\sigma + \psi c_1(\delta + \sigma v)) - \beta_1(\alpha + \delta)(\gamma + \omega + \delta)(\sigma\delta + \delta + \sigma v)] I^* \\
&\quad\quad + \alpha\beta_1\Lambda(\delta + \sigma v) - \delta(\alpha + \delta)(\gamma + \omega + \delta)(\delta + v) = 0\ .
\end{aligned} \tag{29}
$$

Using Vieta's formulas we can determine the sign of the two real solutions of equation (29). Let $I_1^*$ and $I_2^*$ be the solutions of (29) with

$$I_1^* I_2^* = \frac{\alpha\beta_1 \Lambda(\delta + \sigma\nu) - \delta(\alpha+\delta)(\gamma+\omega+\delta)(\delta+\nu)}{\sigma\beta_1^2[\alpha\psi c_1 - (a+\delta)(\gamma+\omega+\delta)]}\,. \quad (30)$$

The denominator of (30) is strictly negative as

$$\alpha\psi c_1 < a\frac{\psi}{\psi+\delta}(\omega+\gamma) < \alpha(\omega+\gamma) < (\alpha+\delta)(\gamma+\omega+\delta). \quad (31)$$

On the other hand, the numerator of (30) is positive when

$$a\beta_1\Lambda(\delta+\sigma\nu) - \delta(\alpha+\delta)(\gamma+\omega+\delta)(\delta+\nu) > 0\ ,$$

thus, according to (15), $R_0 > 1$. Consequently, if $R_0 > 1$, we have solutions with opposite sign, leading to a unique endemic equilibrium, as the negative solution must be rejected according to Theorem 1. Using relations (21) – (26) we calculate the remaining quantities of the endemic equilibrium.

■

Another important issue is to investigate the stability of system (3) examined on the DFE. We prove the local stability of the proposed system based on the value of the basic reproduction number $R_0$, using the formula provided by Theorem 2. More specifically, when $R_0 < 1$, the stability of the model indicates a recursion of system (3) to the initial state when time tends to infinity.

**Theorem 5.** The disease-free equilibrium (DFE) $\boldsymbol{\Upsilon}^0$ is asymptotically stable if and only if $R_0 < 1$, marginally stable when $R_0 = 1$ and unstable otherwise.

**Proof.** Firstly, we remove the equation presenting the flow of the deceased cases in system (3), since $D(t)$ does not participate in the remaining 7 differential equations and can be calculated based on the statement $D(t) = N(t) - \sum(S,E,I,H,C,R,V)(t)$. We linearize (3), by taking the respective Jacobian matrix evaluated on the DFE $\boldsymbol{\Upsilon}^0$,

$$\boldsymbol{J}_{DFE} = \begin{pmatrix} -(\nu+\delta) & 0 & -\beta_1 S^0 & 0 & 0 & \psi & 0 \\ 0 & -(\alpha+\delta) & \beta_1(S^0+\sigma V^0) & 0 & 0 & 0 & 0 \\ 0 & \alpha & -(\gamma+\omega+\delta) & 0 & 0 & 0 & 0 \\ 0 & 0 & \gamma & -(\lambda+\kappa+\mu+\delta) & 0 & 0 & 0 \\ 0 & 0 & 0 & \lambda & -(\tau+\rho+\delta) & 0 & 0 \\ 0 & 0 & \omega & \kappa & \tau & -(\psi+\delta) & 0 \\ \nu & 0 & -\sigma\beta_1 V^0 & 0 & 0 & 0 & -\delta \end{pmatrix}. \quad (32)$$

System (3) is asymptotically stable when all the eigenvalues of the Jacobian $J_{DFE}$ lay on the left side of the complex plane, thus having negative real parts. The characteristic polynomial of $\boldsymbol{J}_{DFE}$ is

$$\begin{aligned} p(s) &= \det(\boldsymbol{J}_{DFE} - s\boldsymbol{I}_{7x7}) \\ &= (s+\delta)(s+\lambda+\kappa+\mu+\delta)(s+\nu+\delta)(s+\tau+\rho+\delta)(s+\delta+\psi) \\ &\qquad (s^2 + (a+\gamma+\omega+2\delta)s + \delta(\alpha+\gamma+\delta+\omega) + \alpha(\gamma+\omega-\beta_1 S^0 - \sigma\beta_1 V^0)) = 0, \end{aligned} \quad (33)$$

leading to 5 negative eigenvalues, namely $s_1 = -\delta, s_2 = -(\lambda+\kappa+\mu+\delta), s_3 = -(\nu+\delta)$, $s_4 = -(\tau+\rho+\delta)$ and $s_5 = -(\delta+\psi)$.

For the quadratic polynomial (33), we apply the second-order Routh-Hurwitz criterion, where the roots of the quadratic polynomial, lay on the left-hand side of the complex plane when coefficients $(a+\gamma+\omega+2\delta)$ and $\delta(\alpha+\gamma+\delta+\omega)+\alpha(\gamma+\omega-\beta_1 S^0-\sigma\beta_1 V^0)$ are both positive. As $(a+\gamma+\omega+2\delta) > 0$, then the Routh-Hurwitz criterion is satisfied if and only if

$$\delta(\alpha+\gamma+\delta+\omega)+\alpha(\gamma+\omega-\beta_1 S^0-\sigma\beta_1 V^0)>0$$

or

$$\beta_1 a(S^0+\sigma V^0)<a(\gamma+\omega+\delta)+\delta(\gamma+\omega+\delta)$$

or

$$\beta_1 a(S^0+\sigma V^0)<(a+\delta)(\gamma+\omega+\delta)$$

and consequently $R_0 < 1$, showing that system (3) is locally asymptotically stable if and only if $R_0 < 1$. In case where $R_0 = 1$, the quadratic polynomial of (33) takes the form

$$s^2+(a+\gamma+\omega+2\delta)s=0\ ,$$

leading to the real solutions

$$s_1=0\,,\qquad s_2=-(a+\gamma+\omega+2\delta)<0. \tag{34}$$

Solution $s_1$ is laying on the imaginary axis $y'y$, characterizing a marginally stable system, as the remaining 6 eigenvalues are laying at the left-hand side of the complex plane. For $R_0 > 1$ the quadratic equation has roots with positive real parts, thus rendering system (3) unstable. This means that the existence of even a small number of infections can lead to a rapid increase of contagiousness. ■

*2.3 State Space Representation of the epidemiological model and Unscented Kalman Filter*

For the transition from the deterministic to the stochastic approach we should reformulate the system of ODEs into a state-space representation. This effort is based on the discretized ODE system (4) for $\Delta t = 1$, as the available observations of the pandemic concerning cases are published daily. Vector $\boldsymbol{x}_k = [S_k\ \ E_k\ \ I_k\ \ H_k\ \ C_k\ \ D_k\ \ V_k]^T$ constitutes the state vector that describes the number of susceptible, exposed, infectious, hospitalized, ICU admitted, recovered, deceased and vaccinated cases at time step $k$. Aiming to provide a dynamic parameter estimation capable of describing the fluctuating dynamics of the pandemic, we augment the state vector $\boldsymbol{x}_k$, by adding the time-varying parameters of the model resulting in the vector $\widetilde{\boldsymbol{x}}_k = [\boldsymbol{x}_k, \boldsymbol{\theta}_k]$, where $\boldsymbol{\theta}_k = [\beta_k, \gamma_k, \lambda_k, \kappa_k, \mu_k, \tau_k, \rho_k]^T$. Hence, the parameter estimation is performed in parallel with the state estimation based on the same iterative procedure.

Thus, the state-space representation takes the form

$$\widetilde{\boldsymbol{x}}_k = f(\widetilde{\boldsymbol{x}}_{k-1})+\boldsymbol{v}_k\ , \tag{35}$$

where

$$f(\widetilde{\boldsymbol{x}}_{k-1})=\begin{bmatrix} \Lambda+\left(1-\left(\frac{\beta_{k-1}I_{k-1}}{N}+\delta+v\right)\right)S_{k-1}+\psi R_{k-1} \\ (1-(a+\delta))E_{k-1}+\frac{\beta_{k-1}S_{k-1}I_{k-1}}{N}+\frac{\sigma\beta_{k-1}V_{k-1}I_{k-1}}{N} \\ (1-(\gamma_{k-1}+\omega+\delta))I_{k-1}+\alpha E_{k-1} \\ \left(1-(\lambda_{k-1}+\kappa_{k-1}+\mu_{k-1}+\delta)\right)H_{k-1}+\gamma_{k-1}I_{k-1} \\ (1-(\tau_{k-1}+\rho_{k-1}+\delta))C_{k-1}+\lambda_{k-1}H_{k-1} \\ (1-(\psi+\delta))R_{k-1}+\omega I_{k-1}+\kappa_{k-1}H_{k-1}+\tau_{k-1}C_{k-1} \\ D_{k-1}+\mu_{k-1}H_{k-1}+\rho_{k-1}C_{k-1} \\ \left(1-\left(\frac{\sigma\beta_{k-1}I_{k-1}}{N}+\delta\right)\right)V_{k-1}+vS_{k-1} \\ \boldsymbol{\theta}_{k-1} \end{bmatrix}, \tag{36}$$

while vector $\boldsymbol{v}_k$ represents the addition of Gaussian noise with covariance matrix $\boldsymbol{Q}$, indicating the transition from the deterministic to the stochastic point of view. The use of white noise encapsulates the uncertainty that accompanies the transitions between the states that the deterministic approach is unable to describe. In addition, the inclusion of noise allows the selection of a low-complexity state-space representation, retaining only the most influential transitions. Transitions with negligible impact during the examined period, such as movements from the infectious to the deceased or the ICU admitted compartments become now part of the Gaussian noise, leading to a computationally economic model, while less time-dependent parameters need to be estimated. Incorporating all insignificant transitions trying to express every extreme case, would lead to a computationally expensive model with an increased possibility of overfitting.

In order to exploit the available observations, we need a second stochastic equation connecting the hidden and observable states of the phenomenon. Equation (37) addresses this necessity, where $\boldsymbol{w}_k$ also represents white Gaussian noise with zero mean and covariance matrix $\boldsymbol{R}$.

$$\boldsymbol{y}_k = h(\widetilde{\boldsymbol{x}}_k) + \boldsymbol{w}_k \,. \tag{37}$$

The addition of noise in equation (37) is also indispensable due to uncertainties associated with daily recordings, mentioned in the introduction. This element also contributes to the adaptation of the stochastic methodology. In our case, $h$ is a linear function and can be represented by a matrix $\boldsymbol{H}$ of dimension 6×15, as

$$\boldsymbol{H} = (\boldsymbol{0}_{6\times 2} \quad \boldsymbol{I}_{6\times 6} \quad \boldsymbol{0}_{6\times 7}). \tag{38}$$

Thus, equation (37) can be reformulated in the form

$$\boldsymbol{y}_k = \boldsymbol{H}\widetilde{\boldsymbol{x}}_k + \boldsymbol{w}_k \,. \tag{39}$$

As we have completed the presentation of the stochastic model, we emphasize the application of Kalman filter methods such as EKF and UKF to our state-space model. The EKF is an extended version of the widely used Kalman Filter, which is ideal for systems that can be approximated using linear dynamics. EKF makes use of the Jacobian matrices that describe both state and observation systems, aiming to locally linearize the non-linear dynamics of the model. Hence, EKF provides adequate estimations in nonlinear problems that do not exhibit a strong nonlinear behavior (Einicke and White 1999).

The UKF is a version of Kalman filter that aims to eliminate the problems caused by the nonlinearities of the system, using a set of sigma points that describe the distribution of states and observations. The purpose of both EKF and UKF is the efficient elimination of the noise that accompanies the available observations, with the aim of revealing the true evolution of the states. Comparing these 2 algorithms provides valuable insights into the suitability of these methodologies for investigating epidemiological phenomena.

Firstly, we assume an initialization of the states and the corresponding covariance matrix with $\overline{\widetilde{\boldsymbol{x}}}_{\boldsymbol{0}} = E[\widetilde{\boldsymbol{x}}_0]$ and $\boldsymbol{P}_0 = E[(\widetilde{\boldsymbol{x}}_0 - \overline{\widetilde{\boldsymbol{x}}}_{\boldsymbol{0}})(\widetilde{\boldsymbol{x}}_0 - \overline{\widetilde{\boldsymbol{x}}}_{\boldsymbol{0}})^T]$. For the implementation of the UKF methodology, we construct a series of sigma points $\boldsymbol{s}^{(i)}$ with corresponding first-order weights $w_i^a$ and second-order weights $w_i^c$. The $2L + 1$ sigma points and weights can be calculated as

$$\boldsymbol{s}^0 = \overline{\widetilde{\boldsymbol{x}}}_{k-1|k-1} \tag{40}$$

$$w_0^a = \frac{a^2 - 1}{a^2} \tag{41}$$

$$w_0^c = w_0^a + 1 - a^2 + \beta \tag{42}$$

$$\boldsymbol{s}^{(i)} = \overline{\widetilde{\boldsymbol{x}}}_{k-1|k-1} + \alpha\sqrt{L}\boldsymbol{A}_i, \;\; i = 1, \dots, L \tag{43}$$

$$\boldsymbol{s}^{(L+i)} = \overline{\widetilde{\boldsymbol{x}}}_{k-1|k-1} - \alpha\sqrt{L}\boldsymbol{A}_i, \;\; i = 1, \dots, L \tag{44}$$

$$w_i^a = w_i^c = \frac{1}{2a^2L}, \;\; i = 1, \dots, 2L \tag{45}$$

(Wan and Van der Merwe 2001). The vector $\boldsymbol{A}_i$ is the $i$-th column of matrix $\boldsymbol{A}$ of $\boldsymbol{P}_{k-1|k-1} = \boldsymbol{A}\boldsymbol{A}^T$, while $\boldsymbol{A}$ can be calculated using the Cholesky decomposition. Parameter $\alpha$ determines the spread of sigma points around the mean state $\bar{\tilde{\boldsymbol{x}}}$ and is usually defined as $\alpha = 0.001$. The parameter $\beta$ is used to determine the prior knowledge of the distribution of $\tilde{\boldsymbol{x}}$ and for Gaussian distributions the optimal value is $\beta = 2$ (Wan and Van der Merwe 2000).

Like other Kalman filter methods, the UKF utilizes an iterative prediction-update algorithm aiming to provide the best possible estimations. During the prediction process the algorithm takes advantage of the sigma points by propagating them through the nonlinear state-space function $\tilde{\boldsymbol{x}}^{(i)} = f\left(\boldsymbol{s}^{(i)}\right), i = 0, \dots, 2L,$ leading to the calculation of a weighted mean and covariance matrix for the $k$-th prediction step,

$$\bar{\tilde{\boldsymbol{x}}}_{k|k-1} = \sum_{i=0}^{2L} w_i^a \tilde{\boldsymbol{x}}^{(i)} \tag{46}$$

$$\boldsymbol{P}_{k|k-1} = \sum_{i=0}^{2L} w_i^c (\tilde{\boldsymbol{x}}^{(i)} - \bar{\tilde{\boldsymbol{x}}}_{k|k-1})\left(\tilde{\boldsymbol{x}}^{(i)} - \bar{\tilde{\boldsymbol{x}}}_{k|k-1}\right)^T + \boldsymbol{Q}\ . \tag{47}$$

Given the mean and error covariance matrix of the $k$-th prediction step, $\bar{\tilde{\boldsymbol{x}}}_{k|k-1}$ and $\boldsymbol{P}_{k|k-1}$, we generate new sigma points $\boldsymbol{s}'^{(i)}$ according to (40), (43) and (44) (Särkkä 2007). These points are propagated in the $k$-th update step through the observation function $\boldsymbol{y}^{(i)} = h\left(\boldsymbol{s}'^{(i)}\right), i = 0, \dots, 2L$. Then, the weighted mean and the covariance of the transformed points are calculated

$$\bar{\boldsymbol{y}} = \sum_{i=0}^{2L} w_i^a \boldsymbol{y}^{(i)} \tag{48}$$

$$\boldsymbol{S}_k = \sum_{i=0}^{2L} w_i^c (\boldsymbol{y}^{(i)} - \bar{\boldsymbol{y}})\left(\boldsymbol{y}^{(i)} - \bar{\boldsymbol{y}}\right)^T + \boldsymbol{R}. \tag{49}$$

In addition, the cross covariance and the Kalman gain matrices can be computed,

$$\boldsymbol{C}_{sz} = \sum_{i=0}^{2L} w_i^c (\boldsymbol{s}^{(i)} - \bar{\tilde{\boldsymbol{x}}}_{k|k-1})\left(\boldsymbol{y}^{(i)} - \bar{\boldsymbol{y}}\right)^T\ , \tag{50}$$

$$\boldsymbol{K}_k = \boldsymbol{C}_{sz}\, \boldsymbol{S}_k^{-1}\ . \tag{51}$$

Finally, the updated mean and error covariance matrix are derived

$$\bar{\tilde{\boldsymbol{x}}}_{k|k} = \bar{\tilde{\boldsymbol{x}}}_{k|k-1} + \boldsymbol{K}_k(\boldsymbol{y}_k - \bar{\boldsymbol{y}})\ , \tag{52}$$

$$\boldsymbol{P}_{k|k} = \boldsymbol{P}_{k|k-1} - \boldsymbol{K}_k \boldsymbol{S}_k \boldsymbol{K}_k^T. \tag{53}$$

The same procedure continues until we reach the final time step $K$. The following tables display the operation of EKF and UKF algorithms with dynamic parameter estimation.

**Algorithm 1.** EKF with dynamic parameter estimation

*# Initialization*

$\bar{\tilde{\boldsymbol{x}}}_{\boldsymbol{0}} = E[\tilde{\boldsymbol{x}}_{\boldsymbol{0}}],\ \boldsymbol{P}_0 = E[(\tilde{\boldsymbol{x}}_{\boldsymbol{0}} - \bar{\tilde{\boldsymbol{x}}}_{\boldsymbol{0}})(\tilde{\boldsymbol{x}}_{\boldsymbol{0}} - \bar{\tilde{\boldsymbol{x}}}_{\boldsymbol{0}})^T$

**for** $k = 1{:}K$ **do**

*# State-Parameter Prediction Step*

$\bar{\tilde{\boldsymbol{x}}}_{k|k-1} = f(\bar{\tilde{\boldsymbol{x}}}_{k-1|k-1})$

$\boldsymbol{P}_{k|k-1} = \boldsymbol{F}_k \boldsymbol{P}_{k|k-1} \boldsymbol{F}_{\boldsymbol{k}}^T + \boldsymbol{Q}$ ; where $\boldsymbol{F}_k = \frac{\partial f}{\partial \boldsymbol{x}}|_{\bar{\tilde{x}}_{k-1|k-1}}$

*# Measurement Update Step*

$\boldsymbol{S}_k = \boldsymbol{H}_k \boldsymbol{P}_{k|k-1} \boldsymbol{H}_k^T + \boldsymbol{R}$ ; where $\boldsymbol{H}_k = \frac{\partial h}{\partial \boldsymbol{x}}|_{\bar{\tilde{x}}_{k|k-1}}$

$\boldsymbol{K}_k = \boldsymbol{P}_{k|k-1} \boldsymbol{H}_k^T \boldsymbol{S}_k^{-1}$

$\bar{\tilde{\boldsymbol{x}}}_{k|k} = \bar{\tilde{\boldsymbol{x}}}_{k|k-1} + \boldsymbol{K}_k(\boldsymbol{y}_k - h(\bar{\tilde{\boldsymbol{x}}}_{k|k-1}))$

$\boldsymbol{P}_{k|k} = (\boldsymbol{I} - \boldsymbol{K}_k \boldsymbol{H}_k)\boldsymbol{P}_{k|k-1}$

**end**

**Algorithm 2.** UKF with dynamic parameter estimation

*# Initialization*

$\bar{\tilde{\boldsymbol{x}}}_{\boldsymbol{0}} = E[\tilde{\boldsymbol{x}}_{\boldsymbol{0}}],\ \boldsymbol{P}_0 = E[(\tilde{\boldsymbol{x}}_{\boldsymbol{0}} - \bar{\tilde{\boldsymbol{x}}}_{\boldsymbol{0}})(\tilde{\boldsymbol{x}}_{\boldsymbol{0}} - \bar{\tilde{\boldsymbol{x}}}_{\boldsymbol{0}})^T$

*# Set the weights of sigma points*

$w_0^a = \frac{a^2 - 1}{a^2}, w_0^c = w_0^a + 1 - a^2 + \beta$

$w_i^a = w_i^c = \frac{1}{2a^2 L},\quad i = 1, \ldots, 2L$

**for** $k = 1{:}K$ **do**

*# Calculate the sigma points*

$\boldsymbol{s}^0 = \bar{\tilde{\boldsymbol{x}}}_{k-1|k-1}$

**for** $i = 1{:}L$ **do**

$\boldsymbol{s}^{(i)} = \bar{\tilde{\boldsymbol{x}}}_{k-1|k-1} + \alpha\sqrt{L}\boldsymbol{A}_i$ ; where $\boldsymbol{A}_i$ is the $i$-th column of matrix $\boldsymbol{A}$ of $\boldsymbol{P}_{k-1|k-1} = \boldsymbol{A}\boldsymbol{A}^T$

$\boldsymbol{s}^{(L+i)} = \bar{\tilde{\boldsymbol{x}}}_{k-1|k-1} - \alpha\sqrt{L}\boldsymbol{A}_i$

**end**

*# State-Parameter Prediction Step*

**for** $i = 0{:}2L$ **do**

$\tilde{\boldsymbol{x}}^{(i)} = f(\boldsymbol{s}^{(i)})$

**end**

$\bar{\tilde{\boldsymbol{x}}}_{k|k-1} = \sum_{i=0}^{2L} w_i^a \tilde{\boldsymbol{x}}^{(i)}$

$\boldsymbol{P}_{k|k-1} = \sum_{i=0}^{2L} w_i^c (\tilde{\boldsymbol{x}}^{(i)} - \bar{\tilde{\boldsymbol{x}}}_{k|k-1})(\tilde{\boldsymbol{x}}^{(i)} - \bar{\tilde{\boldsymbol{x}}}_{k|k-1})^T + \boldsymbol{Q}$

*# Measurement Update Step*

*# Calculate new set of sigma points $\boldsymbol{s}'^{(i)}$ given the prediction estimates $\bar{\tilde{\boldsymbol{x}}}_{k|k-1}$ and $\boldsymbol{P}_{k|k-1}$.*

**for** $i = 0{:}2L$ **do**

$\boldsymbol{y}^{(i)} = h(\boldsymbol{s}'^{(i)})$

**end**

$\bar{\boldsymbol{y}} = \sum_{i=0}^{2L} w_i^a \boldsymbol{y}^{(i)}$

$\boldsymbol{S}_k = \sum_{i=0}^{2L} w_i^c (\boldsymbol{y}^{(i)} - \bar{\boldsymbol{y}})(\boldsymbol{y}^{(i)} - \bar{\boldsymbol{y}})^T + \boldsymbol{R}$

$\boldsymbol{C}_{sz} = \sum_{i=0}^{2L} w_i^c (\boldsymbol{s}^{(i)} - \bar{\tilde{\boldsymbol{x}}}_{k|k-1})(\boldsymbol{y}^{(i)} - \bar{\boldsymbol{y}})^T$

$\boldsymbol{K}_k = \boldsymbol{C}_{sz} \boldsymbol{S}_k^{-1}$

$\bar{\bar{\boldsymbol{x}}}_{k|k} = \bar{\bar{\boldsymbol{x}}}_{k|k-1} + \boldsymbol{K}_k(\boldsymbol{y}_k - \bar{\boldsymbol{y}})$ ; updated mean
$\boldsymbol{P}_{k|k} = \boldsymbol{P}_{k|k-1} - \boldsymbol{K}_k \boldsymbol{S}_k \boldsymbol{K}_k^T$ ; updated covariance matrix

**end**

## 3. Results

### *3.1. Simulation of the deterministic SEIHCRDV model*

In this section, we explore the dynamics of system (3) using the numerical method of 4th-order Runge-Kutta, which is a widely used numerical method for solving ODEs systems. For the purposes of this analysis, we treat the model's parameters as fixed. Firstly, let $\sigma = 0.05$, considering that the average protection rate from the vaccination is 95% (Polack et al. 2020). Moreover, let $a = \frac{2}{11}$ (Evensen et al. 2021), $\psi = \frac{1}{180}$ (Schuler et al. 2021) and $\omega = \frac{1}{14}$ (Katul et al. 2020). The median value of daily new vaccinated cases in France is about 140.000 people, leading to a vaccination rate of $v \approx 0.002$, since the country's total population is about 65.45 million.

Birth-migration rate $\Lambda$ and mortality rate $\delta$ due to non-COVID causes can be assumed to be constant for short periods of time (Zamir et al. 2021; Nadeem et al. 2019). We base the selection of $\Lambda$ and $\delta$ on official data of France (Statista 2021), to be consistent with the application to real data. The mortality rate during 2017-2019 (before the onset of COVID) was stable, showing approximately 9.1 deaths per 1000 people. Therefore, the annual mortality rate from non-COVID sources can be $0.0091$, leading to a daily mortality rate of $\delta = 2.05 \times 10^{-5}$. On the other hand, the total number of births during 2021, was about 10.9 per 1000 people (0.0109 annual birth rate and 2.99$\times 10^{-5}$ daily rate), while the total number of net migration was about 140.000 people, leading to a daily net migration rate of 5.48$\times 10^{-7}$. As a result, we choose $\Lambda = 3.045 \times 10^{-5}\ N$, where $N$ denotes the country's total population.

As the average time intervals of the transitions from the hospitalized to the recovered or the deceased state are 5 and 16 days respectively, we set $\kappa = \frac{1}{5}$ and $\mu = \frac{1}{16}$ (Evensen et al. 2021). Similarly, the mean intervals between an ICU admission and the transition to the recovered or the deceased state are 8 and 15 days, respectively (Roquetaillade et al. 2021). We assume the average transition rate between the hospitalized and ICU admitted states to be 0.1. As a result we set $\tau = \frac{1}{8}$, $\rho = \frac{1}{15}$, while $\lambda = 0.1$. Based on Thompson et al. (2020) we can consider the mean duration of a transition from the infectious to the hospitalized state to be 6.5 days, giving a $\gamma = \frac{2}{13}$. In addition, Faes et al. (2020) report that in a study concerning 14.618 hospitalized patients in Belgium, the mean duration between the symptom onset and hospitalization was 3 to 10.4 days, validating this choice for $\gamma$. Finally, we consider the infection rate fixed with $\beta = 1$.

The following simulated results are obtained utilizing the model provided by system (3). In figure 2, we present the evolution of a pandemic according to the proposed SEIHCRDV model using the 4th-order Runge-Kutta method.

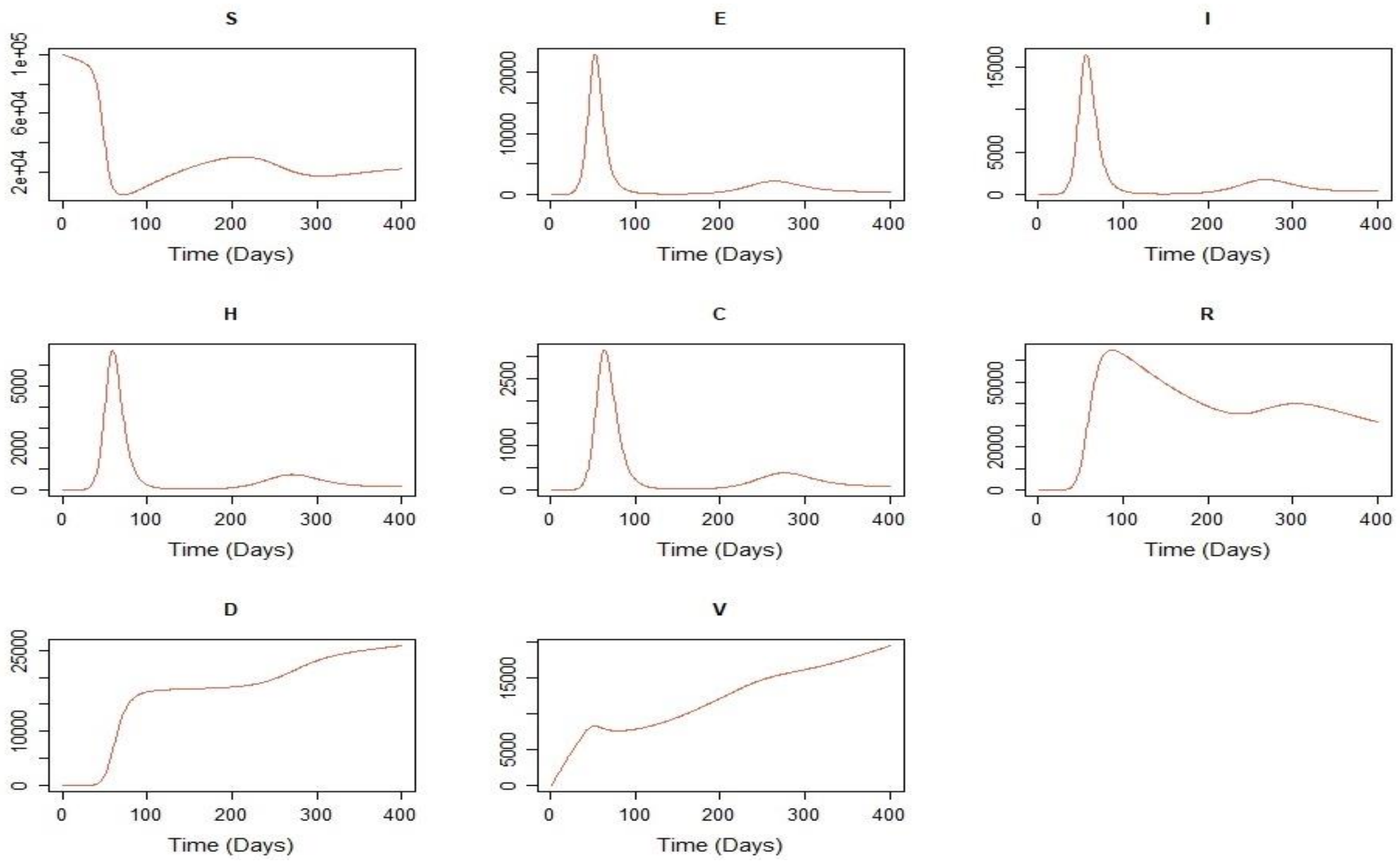

**Figure 2.** Numerical solution of the proposed SEIHCRDV model using 4th-order Runge-Kutta algorithm

We run the simulation for 400 days, using the abovementioned constant parameters. This simulation produces two waves of infection, with the first, most prevalent one reaching its peak after 56 days and the second less severe one after 268 days. Assuming fixed parameters, we restrict our model into providing a pandemic scenario, which does not consider external measures, such as masks or lockdowns that suppress the spread of the disease. In addition, this assumption excludes cases like the emergence of COVID-19 variants, which may affect the rate of hospital or ICU admissions, as well as the death and recovery rates of infectious individuals.

The hospital and ICU admissions curves imitate the infection curve with a relatively short delay of 3 and 8 days, respectively, during the first most prevalent infection wave. We find similar behavior during the second wave. The cumulative recovery curve reaches its first maximum after 87 days, following the upward trend of the infection curve. The displayed decrease until day 238, is due to the transition from the recovered to the susceptible state, as we consider that the antibodies produced by infection last up to 6 months (180 days) on average. We also notice that the curve of deceased cases shows an increasing trend following the emergence of the two infection waves, displaying a much more intense increase during the first infection wave, as expected. Finally, the vaccination curve follows a rarely linear increasing trend, since we refer to the cumulative number of fully vaccinated individuals.

### *3.2 Impact of vaccination*

An important question to investigate is how the various vaccination rates affect the evolution of the pandemic. In this paragraph, we variate the parameter $v$ corresponding to the vaccination rate, over a time interval of 400 days of pandemic. In figure 3, we observe the fluctuations in the infectious cases, the hospital and ICU admissions and the cumulative number of deaths from day 1 to 400. In each graph, there are four curves produced according to vaccination rates $v = 0.002$, $v = 0.003$, $v = 0.004$ and $v = 0.005$, representing an increasing vaccination tendency.

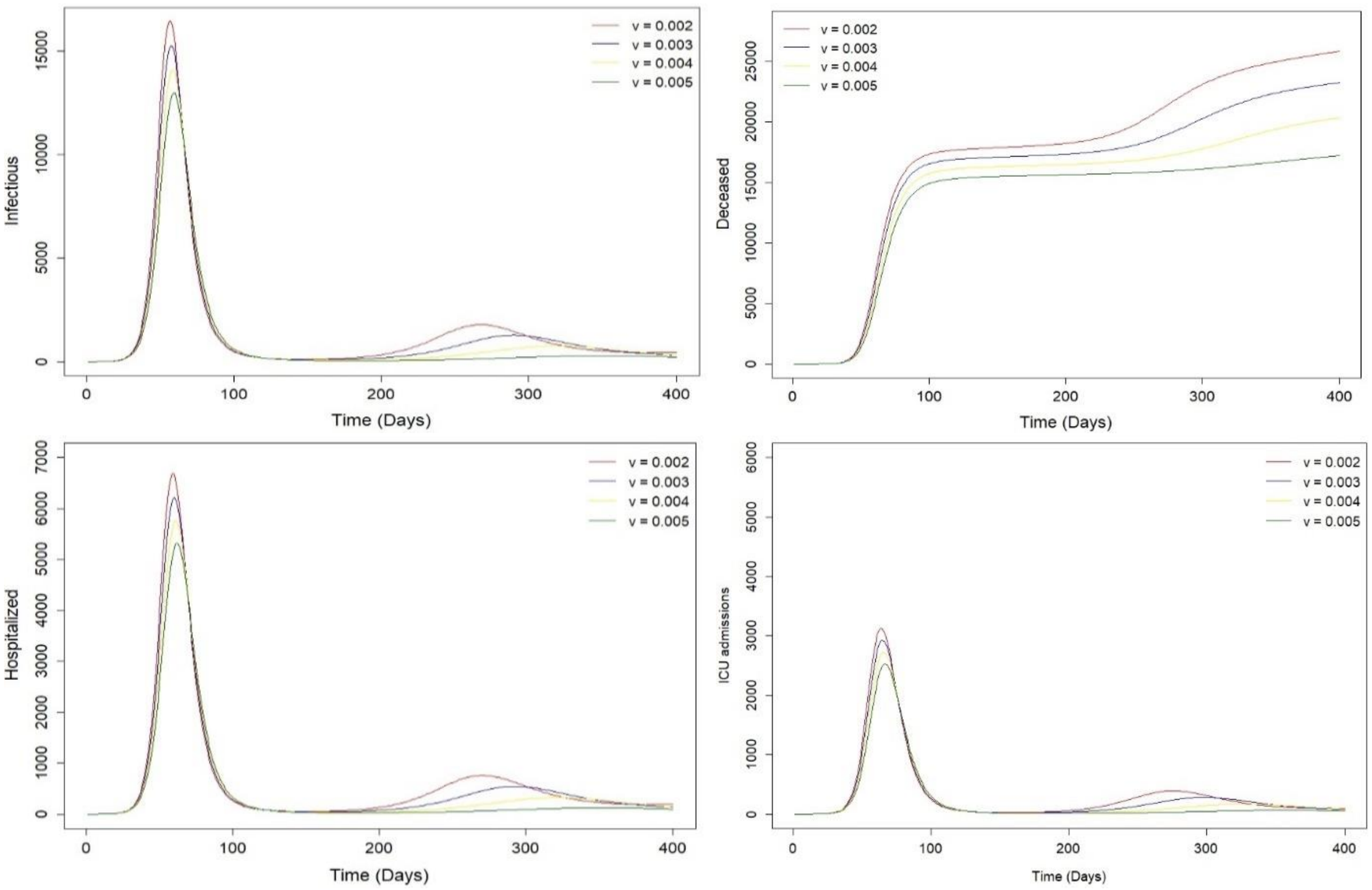

**Figure 3.** Curves of infectious, deceased, hospitalized and ICU admitted cases for increased vaccination rates

By increasing the vaccination rate from $v = 0.002$ to $v = 0.005$, the number of infectious cases, hospitalizations, ICU admissions and deceased individuals, falls gradually revealing an inversely proportional relation between the increase in the number of fully vaccinated individuals and the spread of the pandemic through the affected population. The curves of hospitalizations and ICU admissions follow the behavior of the curve of infectious cases, a phenomenon that is highly expected, as the fewer the infectious cases, the less the hospital and ICU admissions.

An important observation is that the usage of a vaccination rate of $v = 0.005$, not only mitigates the first wave of COVID-19 spread, but almost extinguishes the second wave of the pandemic leading to a stabilization of the disease. Theoretically, this significant deterioration of the pandemic spread could be achieved through a daily relative raise of 1.5% in fully vaccinated cases, confirming the importance of implementing a rapid vaccination process.

*3.3 Impact of an increase in hospital and ICU admissions*

Following the methodology mentioned in the previous paragraph, we variate the parameter $\lambda$ that represents the transition rate from the state of hospitalized cases to ICU state, as long as the rate of hospitalizations $\gamma$. These rates may be affected by the prevalence of COVID's variants like B.1.1.7, B.1.351 (Veneti et al. 2021) or B.1.617.2 (Kidd et al. 2021). More specifically, we choose a $\lambda = 0.1$ for the creation of the initial timeseries of ICU admissions. Hence, three more curves are constructed by increasing the average transition rate by 0.1. Concerning the hospitalizations rate, the initial simulated timeseries is constructed using $\gamma = \frac{2}{13}$ derived from the average time interval from the emergence of symptoms to hospital admission. We aim to examine the pandemic evolution, when this average time interval decreases from 6.5 to 6, 5.5 and 5 days respectively. This procedure may help us examine the differentiations influenced by a highly realistic alteration, as there are many COVID-19 variants that significantly deteriorate the health of the infected individuals, leading to intensified hospital and ICU admission rates. In figure 4, we notice the differentiations concerning the hospitalized, ICU admitted, recovered and deceased cases.

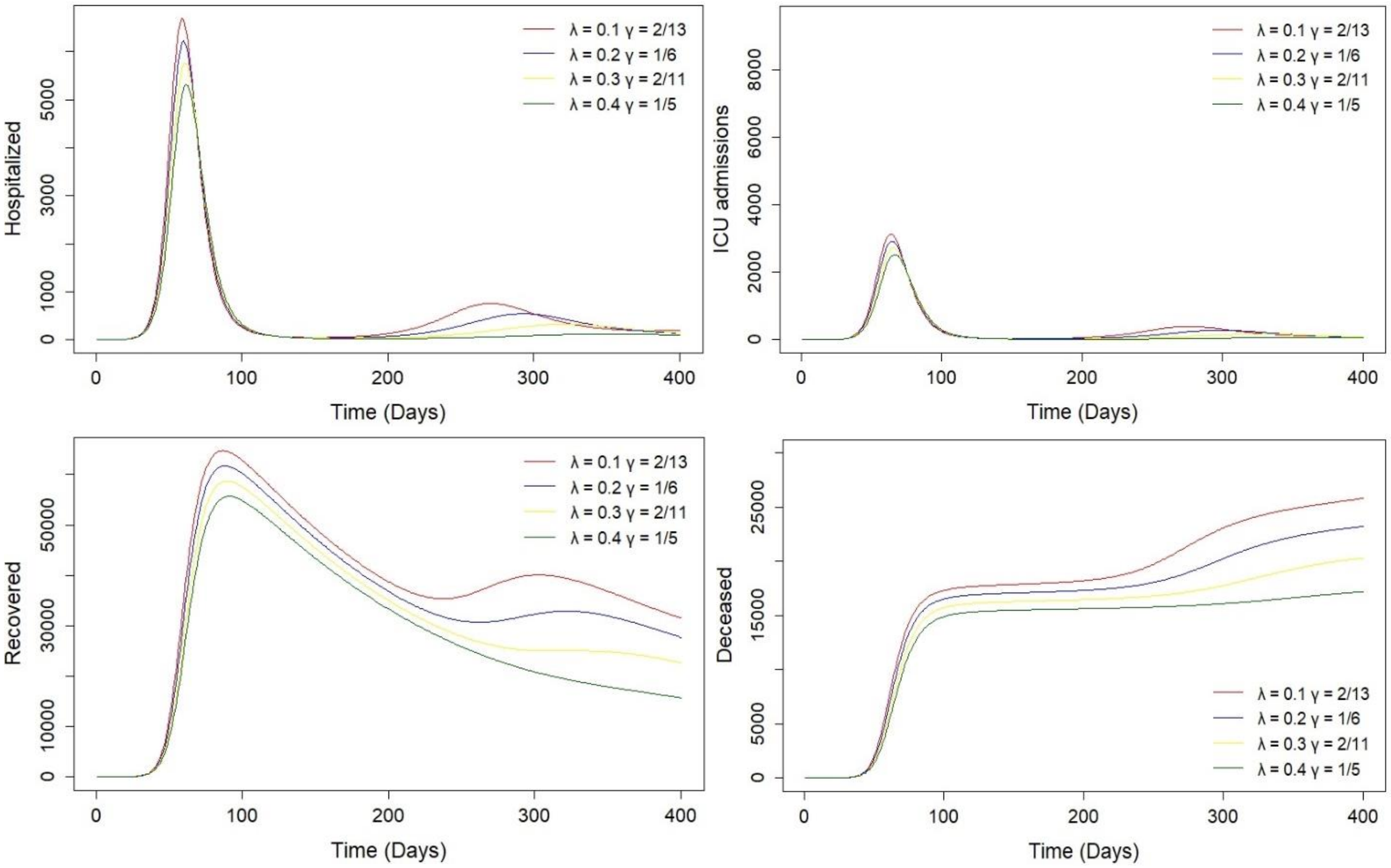

**Figure 4.** Curves of infectious, ICU admitted, recovered and deceased cases for increased hospital and ICU admission rates

According to figure 4, we observe a descending tendency of hospitalized cases. This fact is expected, as we increased the transition rate from the hospitalized to ICU admitted state $\lambda$ more rapidly than the respective rate of hospitalizations $\gamma$. Furthermore, the higher levels of hospital and ICU admissions seem to result in significantly lower cumulative recovered cases, while the cumulative deceased cases reveal an inversely proportional behavior in comparison with the equivalent recovered ones.

*3.4 Resistance of antibodies produced by infection*

As mentioned earlier, the retention period of antibodies produced by the infection varies greatly between different individuals and is highly affected by the subject's immune system as well as the COVID-19 variant that infected them. Therefore, we investigate the impact of the re-sensitization period on the spread of the disease in the population. Figure 5 shows the evolution of infectious, recovered, deceased and vaccinated cases over a re-sensitization period of 3, 6, 9 and 12 months. According to the graphs, we conclude that smaller $\psi$ values result into a delay in the emergence of the second wave, while the number of recovered cases increases continuously. The resulting outcome concerning the evolution of vaccinated cases is interesting, as longer maintenance of antibodies seems to lead to reduced vaccination levels. We may deem that longer re-sensitization intervals moderate the desire of people to get vaccinated as they feel protected by the antibodies of their immune system. However, the need for vaccination reappears for each of the examined 4 re-susceptibility rates, shortly after antibody levels decline.

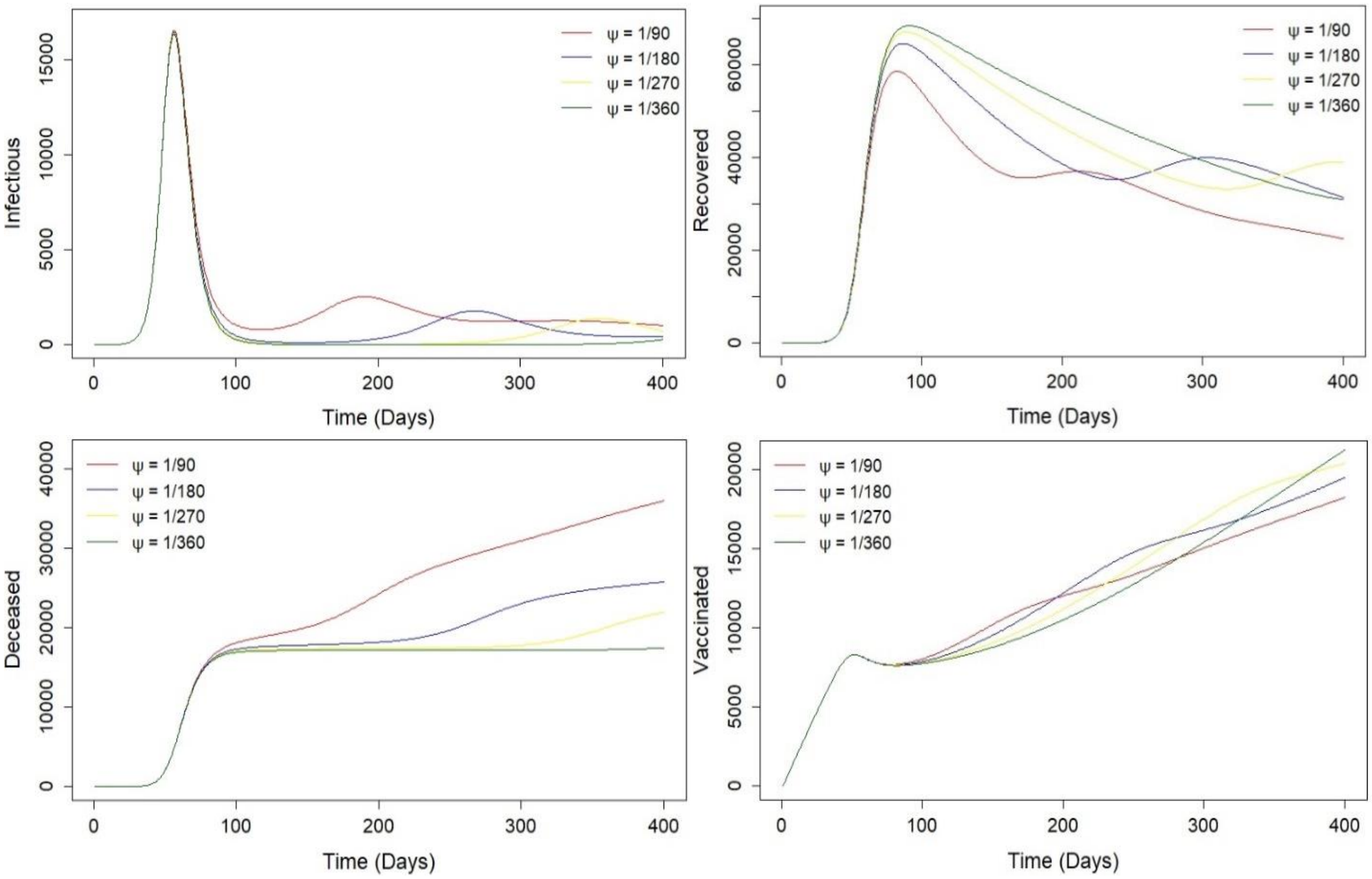

**Figure 5.** The impact of infection-induced antibody resistance

*3.5 Comparison of UKF and EKF efficiency on simulated data*

During this part of the analysis, we investigate the predictive efficiency of the EKF and UKF algorithms based on the aforementioned simulated data using the 4th-order Runge-Kutta numerical solution. The purpose of this section is the identification of the actual states of our model at each time step, during a time interval of 400 days, when we add to the original data a Gaussian noise providing a stochastic equivalent of the proposed deterministic SEIHCRDV model.

We produce the necessary observations for the EKF and UKF models, by adding white noise of zero mean and standard deviation equal to the 10% of each original timeseries' standard deviation, to the six observable timeseries produced by the Runge-Kutta method. We do not examine the performance of UKF and EKF in the susceptible or exposed cases, as these cases are usually considered as hidden states for the system, in most articles dealing with combined epidemiological state-space models.

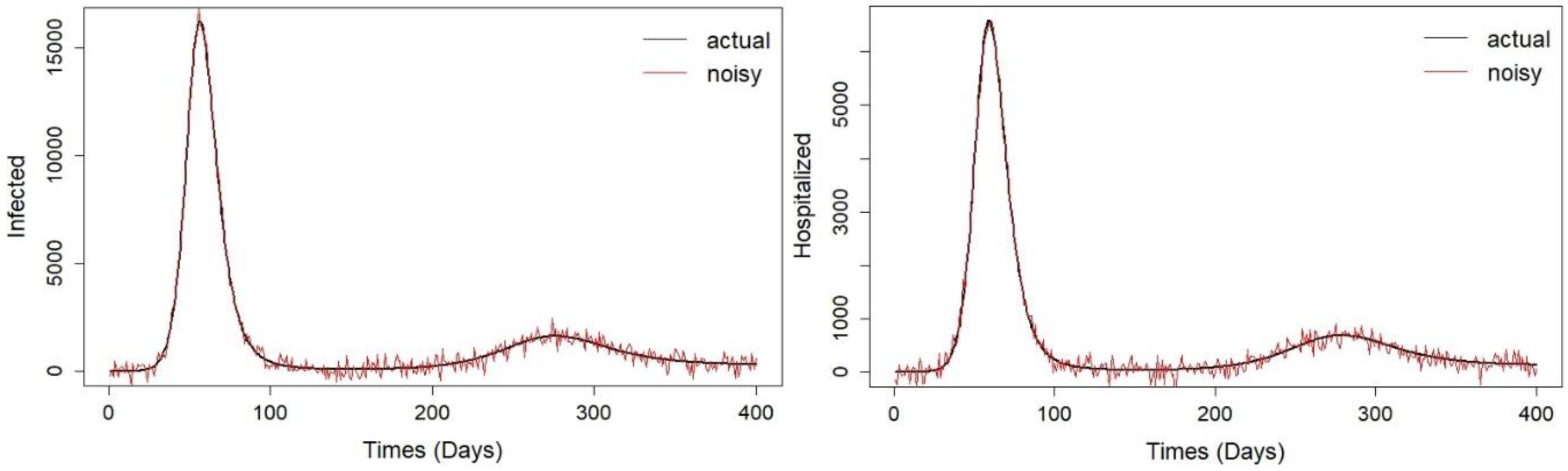

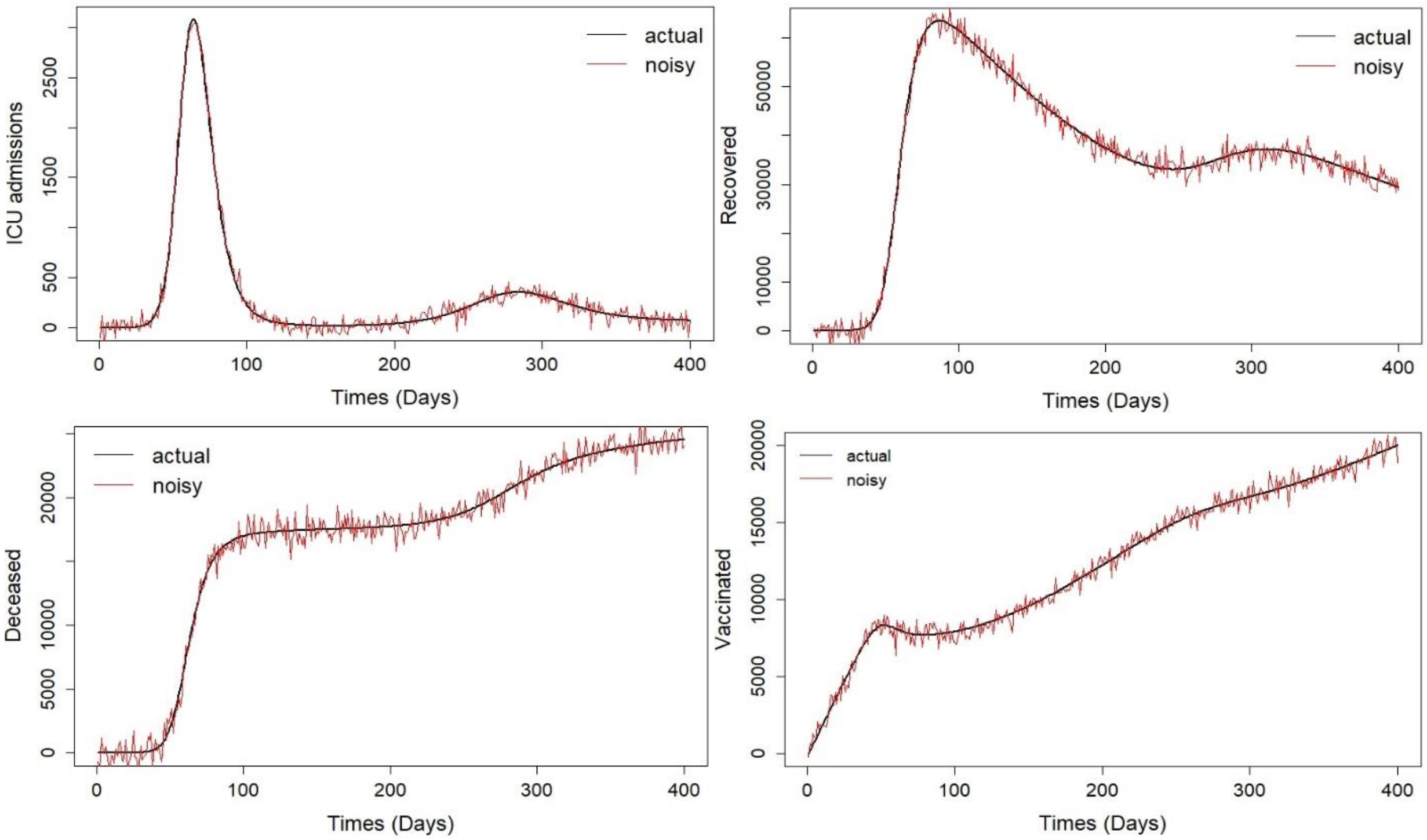

**Figure 6.** Actual simulated data produced by the 4th-order Runge-Kutta method for each of the six model's observable states and the respective time series after the addition of the Gaussian noise.

The purpose of using EKF or UKF, is to efficiently eliminate the effect of noise that accompanies the above observations, to reveal the evolution of the true states; in addition, the comparison of these 2 methodologies will give us valuable inferences about their appropriateness in examining epidemiological models. In figure 6, we observe the actual observations of the epidemic and the same observations after the addition of white noise.

The second goal of this section is to compare the efficiency of the Extended and the Unscented Kalman Filter on simulated data. We use for both algorithms the same noise covariance matrices $\boldsymbol{Q}$ and $\boldsymbol{R}$, as well as the same initial state vectors, $\boldsymbol{x}_0$, and the corresponding initial covariance matrix $\boldsymbol{\Sigma}_0$, to ensure the comparability of the resulting estimations. In figure 7, we can visually compare the estimation capability of EKF and UKF on the simulated pandemic data. In Table 2 we present the RMSE values between the six examined observable timeseries produced by the proposed SEIHCRDV model, the noisy equivalent of these timeseries and the produced estimations of EKF and UKF. In all six cases, the filtering efficacy of UKF overcomes the EKF's, while the widest difference is noticed for the recovered individuals. The EKF provides a greater overestimation of actual infectious, hospitalized and ICU admitted cases compared to the UKF. This can be seen clearly especially in the infection peaks.

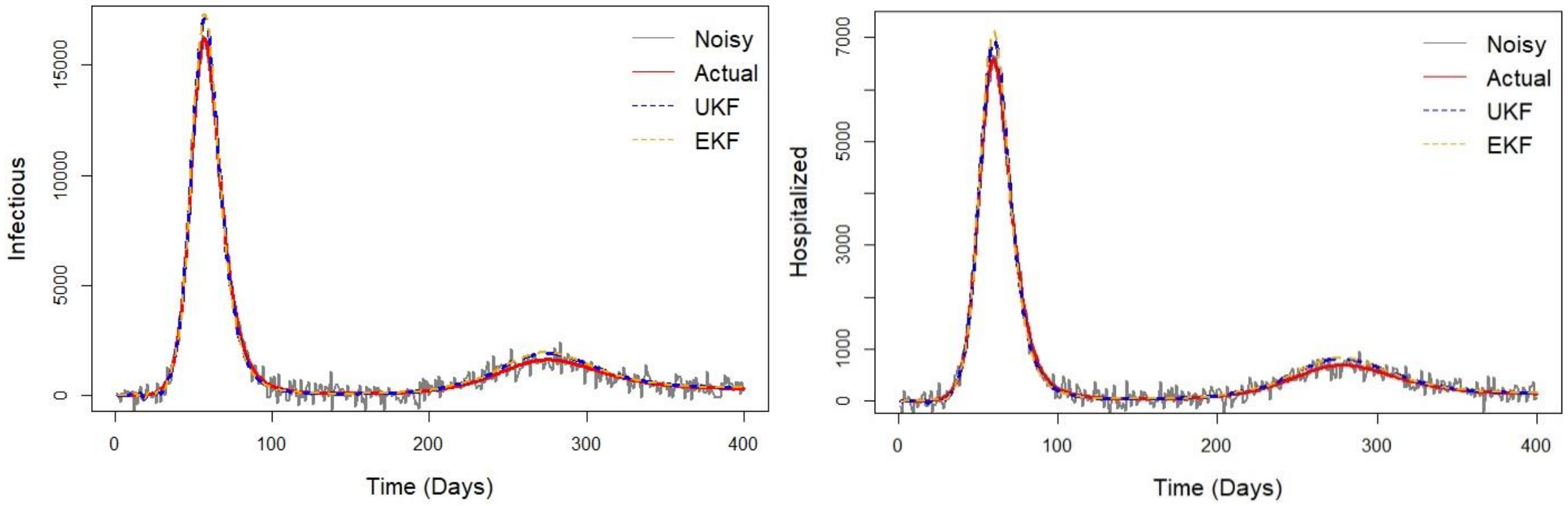

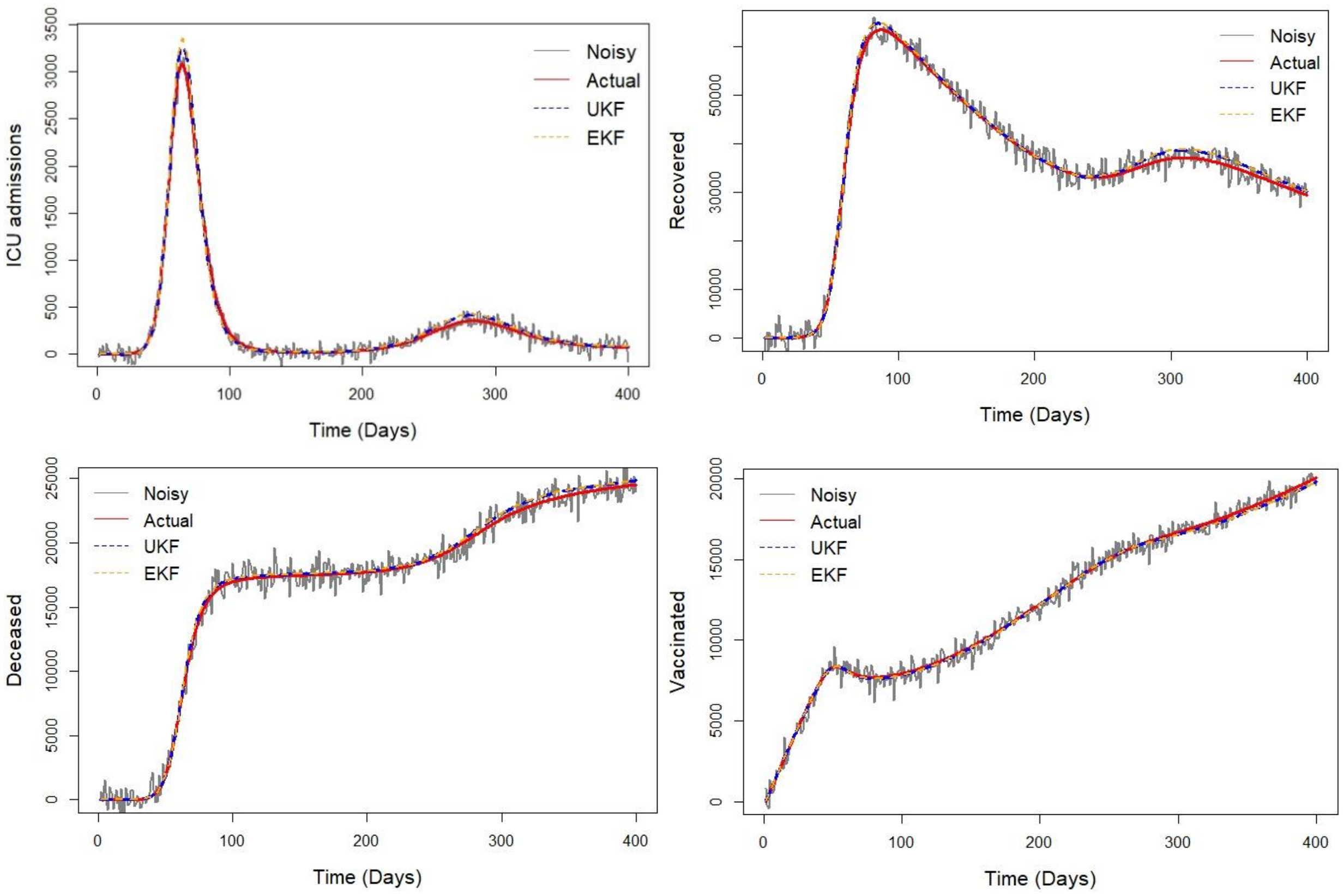

**Figure 7**. Examining the filtering efficacy of EKF and UKF algorithms in simulated noisy data

**Table 2.** RMSEs between noisy data, EKF, UKF estimations and actual simulated data produced by the numerical solution of the SEIHCRDV model

| | Infectious | Hospitalized | ICU | Recovered | Deceased | Vaccinated |
|---|---|---|---|---|---|---|
| Noisy Data | 328.50 | 131.33 | 62.20 | 1792.50 | 752.21 | 495.15 |
| EKF | 224.65 | 99.91 | 51.65 | 918.74 | 271.92 | 126.15 |
| UKF | 194.63 | 83.74 | 42.87 | 847.39 | 253.74 | 117.08 |

### *3.6 Evaluation of COVID-19 pandemic in France based on UKF with dynamic parameter estimation*

This section highlights the fitting and predictive capacity of the dynamic UKF algorithm based on the proposed SEIHCRDV model, applied to daily COVID-19 observations in France. We explore the evolution of the pandemic in France from January 16 to October 7, 2021, providing a time span of 265 days. The fully vaccinated cases represent individuals who have received two doses of Pfizer or Astra Zeneca or one dose of Johnson & Johnson or Moderna vaccines. The observational data used, including infectious (active), hospitalized, ICU admitted, recovered, deceased and vaccinated cases, are collected from the datasets contained on the data.europa.eu webpage. This webpage contains official, daily COVID-19 observations for all European Union members collected from national resources.

The examined period reveals two infection waves that reach their peak around day 80 (April 5) and 210 (August 18), while they are accompanied by respective waves of hospital and ICU admissions. The downward trend of the first infection wave may have been strongly influenced by the daily lockdowns imposed in 16 departments in France, while curfew hours were enforced across the country between 7 p.m. and 6 a.m. every day (March 20). The onset of the second wave around the first week of July (day 170) is highly related to the relaxation of the restrictive measures such as the re-opening of restaurants/bars/cafes with 50% capacity (June 9), the outdoor relaxation of mask-wearing (June 17) and the lifting of the night traffic ban (June 20). Another important factor in the second spread of COVID-19 in France is the emergence of the delta variant in the country, where according to G. Attal (2021) (France 24), the highly contagious delta variant accounts for 40% of new COVID infections.

Based on figure 8, in addition to the first wave examined, the second wave of infection is also prevalent, while the corresponding waves of admissions to hospitals and ICUs are significantly attenuated. This phenomenon shows the improvement of the French medical system against COVID-19, as the experience gained from previous waves of infection helped the experts to understand more details about the characteristics of the disease, leading to more targeted and effective treatment techniques. In addition, the constantly increasing number of fully vaccinated cases plays a key role in the aforementioned phenomenon since it reduces the likelihood of severe infections, leading to fewer hospitalizations and ICU admissions. At the same time, we observe an increase in the total number of deaths but with a strongly declining trend, while the recovery rates increase in parallel with the onset of the two examined infection waves, indicating the more effective treatment of COVID-19 in new cases.

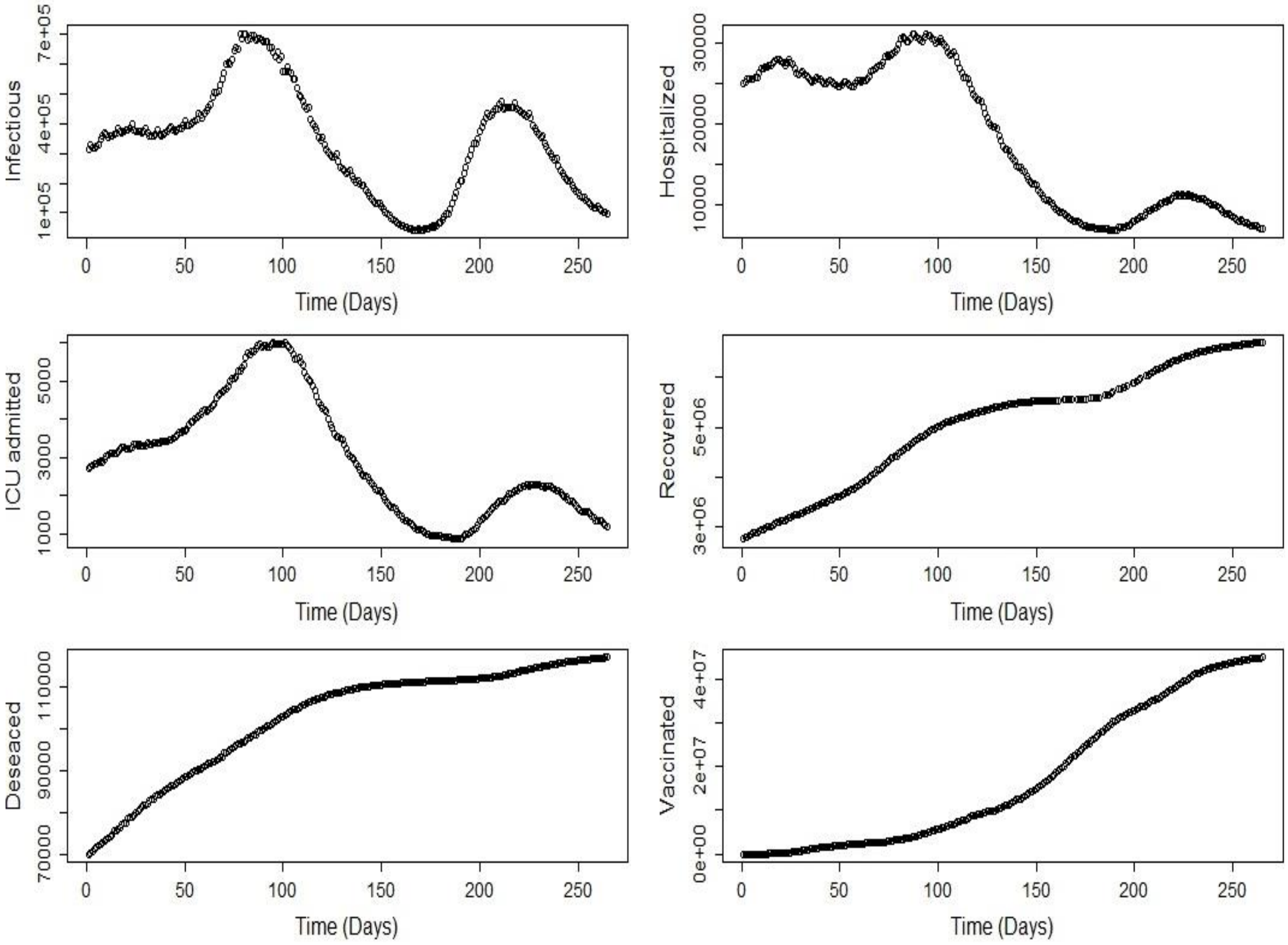

**Figure 8.** The evolution of infectious, hospitalized, ICU admissions, recovered, deceased and vaccinated cases during COVID-19 pandemic in France

Following the first conclusions drawn from the visual examination of figure 8, we next investigate the fitting and predictive efficiency of the UKF applied to the daily COVID-19 observations in France. Specifically, we perform a comparative analysis between an UKF algorithm based on the SEIHCRDV model with constant parameters, a dynamic EKF – SEIHCRDV and a dynamic UKF – SEIHCRDV model with 7 varying parameters. The latter aims to describe more efficiently the dynamics characterizing the spread of COVID in France after the initialization of vaccinations. In figure 9, we present the fitting capacity of UKF and UKF with dynamic parameter estimation using the timeseries of infectious (active), hospitalized, ICU admitted, recovered, deceased and vaccinated cases in France. The most characteristic difference is that the standard UKF – with constant parameters – displays an underestimation of infectious cases especially during the two infection waves in April and August and a more pronounced overestimation of the corresponding number of hospital and ICU admissions, compared to the results derived from the dynamic UKF – SEIHCRDV model.

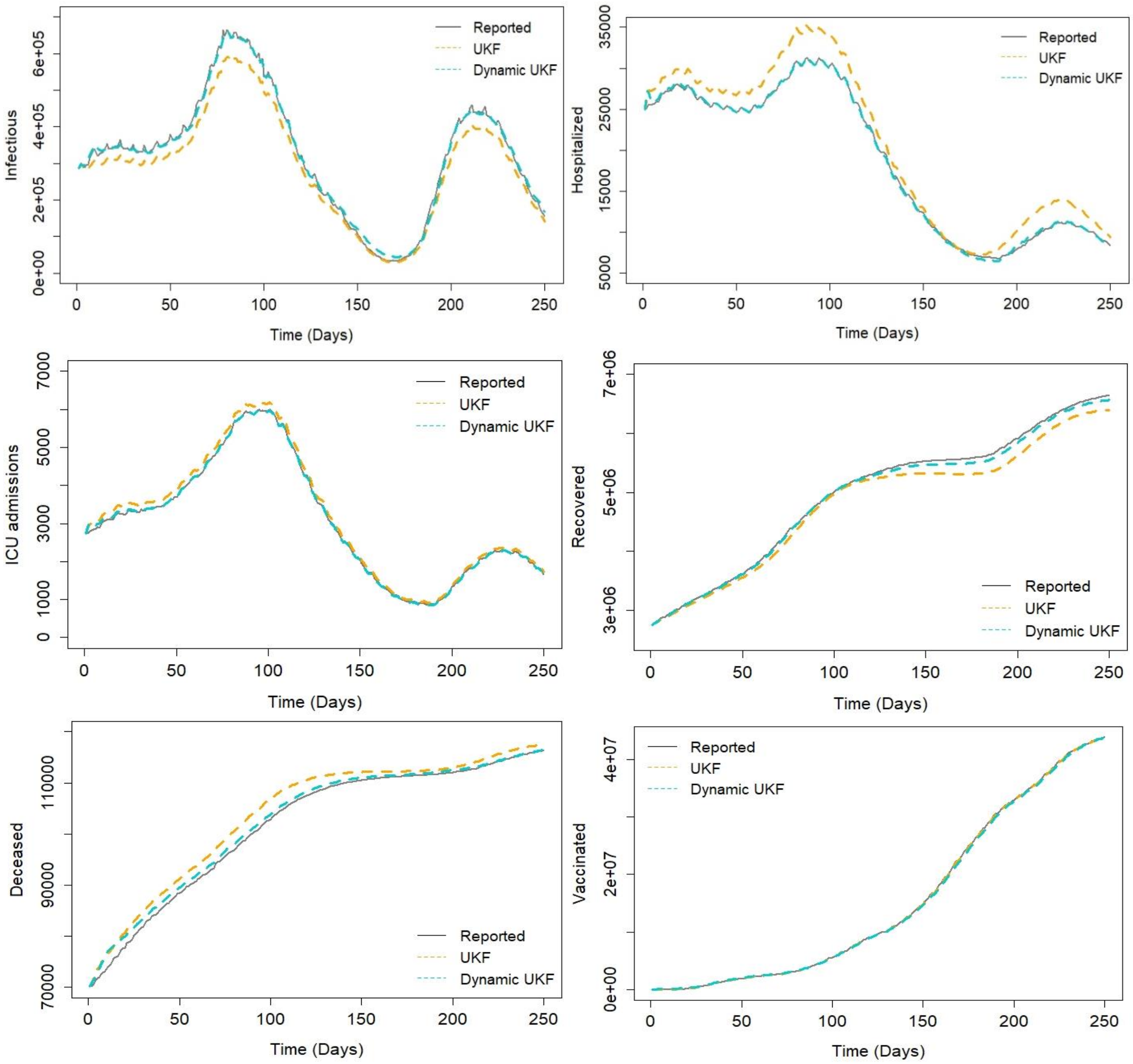

**Figure 9.** Performance comparison of UKF and UKF with dynamic parameter estimation based on the proposed SEIHCRDV model for COVID-19 in France

Table 3 below, displays the respective normalized root mean squared errors (NRMSEs) of the numerical solution of the deterministic SEIHCRDV and the three abovementioned stochastic equivalents of UKF, EKF and UKF with dynamic parameter estimation. The presented NRMSE values are calculated according to the formula presented in (Papageorgiou and Tsaklidis 2021; Pal 2016). Both UKF, dynamic EKF and UKF significantly outperform the deterministic SEIHCRDV equivalent. Furthermore, the proposed UKF with dynamic parameter estimation provides the most accurate description of the pandemic evolution, rarely in all 6 observable states. In particular, the standard UKF model produces an increase in NRMSE values of 309.09%, 765.09%, 361.59%, 261.27% and 149.79% compared to the dynamic UKF – SEIHCRDV, while the dynamic EKF provides an increase of 56.74%, 80.95%, 20.83%, 48.33% and 40.19% for the infectious (active), hospitalized, ICU admitted, recovered and deceased cases respectively.

**Table 3.** NRMSEs of the deterministic SEIHCRDV, UKF, and UKF with dynamic parameter estimation applied on COVID-19 data of France

| | Infectious | Hospitalized | ICU | Recovered | Deceased | Vaccinated |
|---|---|---|---|---|---|---|
| Numerical | 0.962637 | 0.976639 | 0.996749 | 0.989558 | 0.998393 | 0.957881 |
| UKF | 0.226715 | 0.250055 | 0.089036 | 0.143825 | 0.159559 | 0.066471 |
| Dynamic EKF-SEIRD | 0.119928 | - | - | 0.056008 | 0.266164 | - |
| Dynamic EKF-SEIHCRDV | 0.086864 | 0.052304 | 0.042115 | 0.059055 | 0.089552 | 0.018929 |
| **Dynamic UKF-SEIHCRDV** | **0.055420** | **0.028905** | **0.019289** | **0.039811** | **0.063877** | **0.018131** |

An important characteristic that verifies the robustness of the proposed UKF – SEIHCRDV model is the monitoring of the parameter updates (Figure 10). First, the infection rate $\beta$, follows the existence of the two waves of infection that we observe within the period of 250 days. The transition parameters $\gamma$ and $\lambda$ display a relatively decreasing behavior during the evolution of the pandemic, which also seems to be in agreement with the observable timeseries. This phenomenon underlines the slightly reduced risk of severe infection as we move deeper into the vaccination period.

Another notable observation is the descending trend of mortality rates $\mu$ and $\rho$ over the 250 days examined, where the mortality rate of hospitalized cases $\mu$ becomes lower than the mortality rate $\rho$ of ICU admissions. This phenomenon is quite expected, as the total number of deaths increases with a decreasing trend (Figure 8) and the probability of mortality of ICU cases is comparatively higher than the respective mortality of hospitalized cases. Finally, the behavior of the recovery rates $\kappa$ and $\tau$ validates the trustworthiness of our model, as both rates increase significantly over time, emphasizing the impact of vaccination and the adaptability of the France's national medical system to the challenges posed by this new virus. Therefore, the recovery rate $\kappa$ of hospitalized patients displays an ascending trend while continuously reaching higher levels during the 250 days compared to the ICU recovery rate $\tau$. In figure 11, we display the evolution of the reproduction number, during the examined period of 250 days.

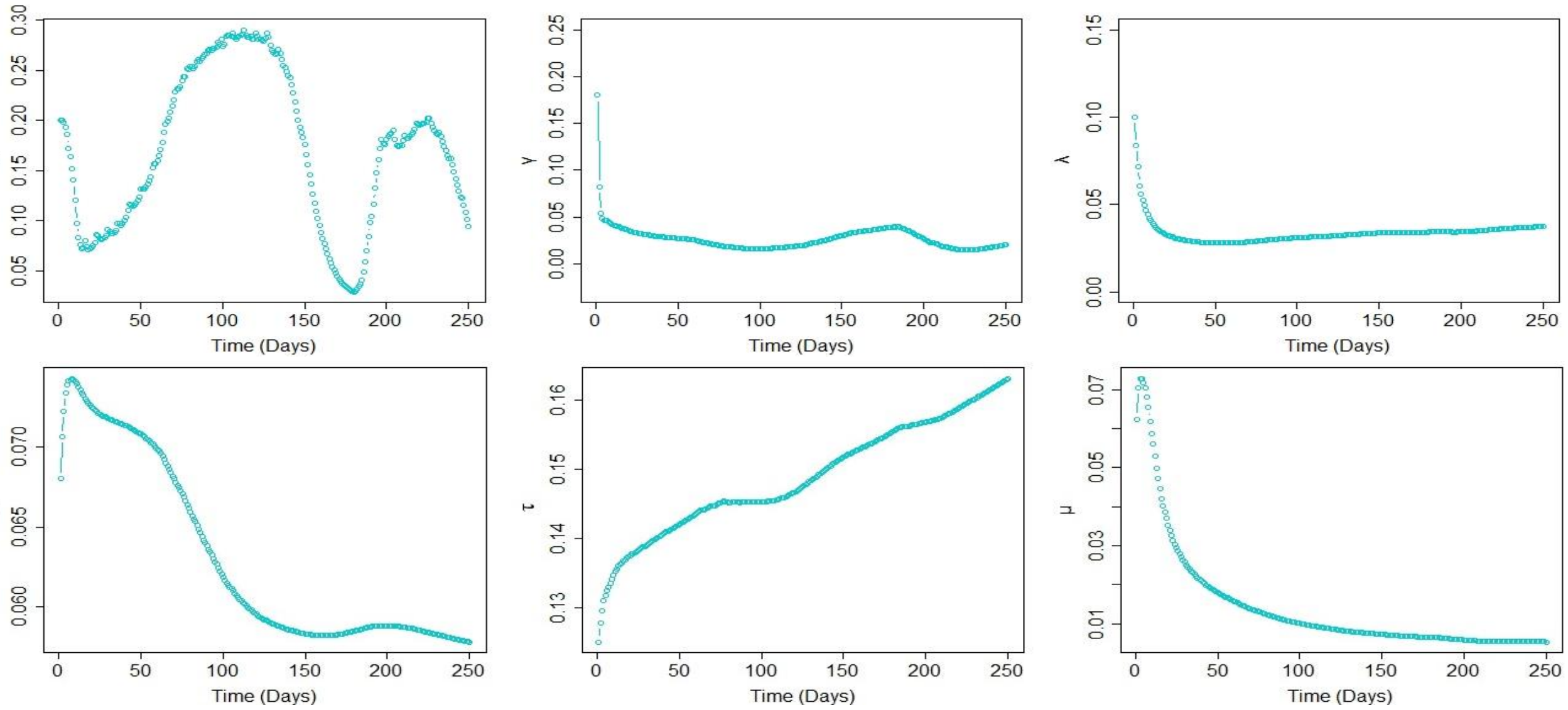

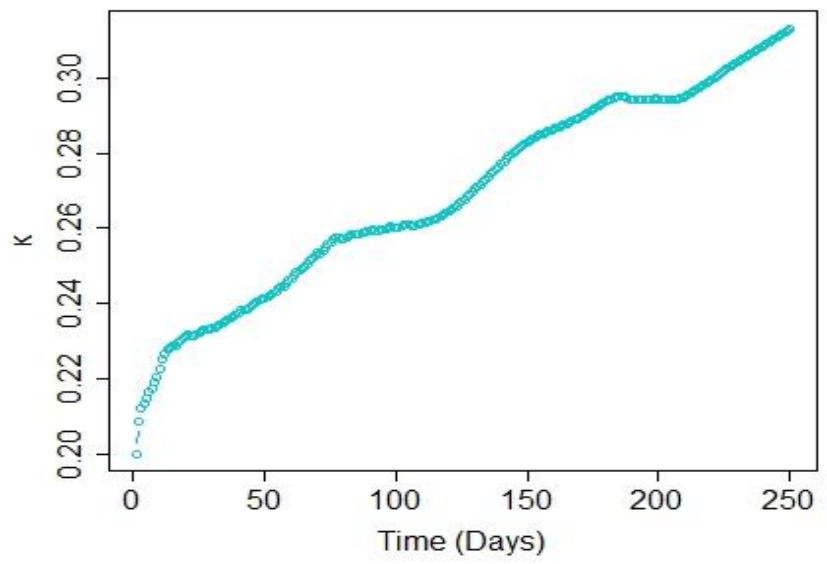

**Figure 10.** Evolution of the 7 non-constant parameters of the SEIHCRDV model, that derive from the application of the dynamic UKF algorithm to daily data of France.

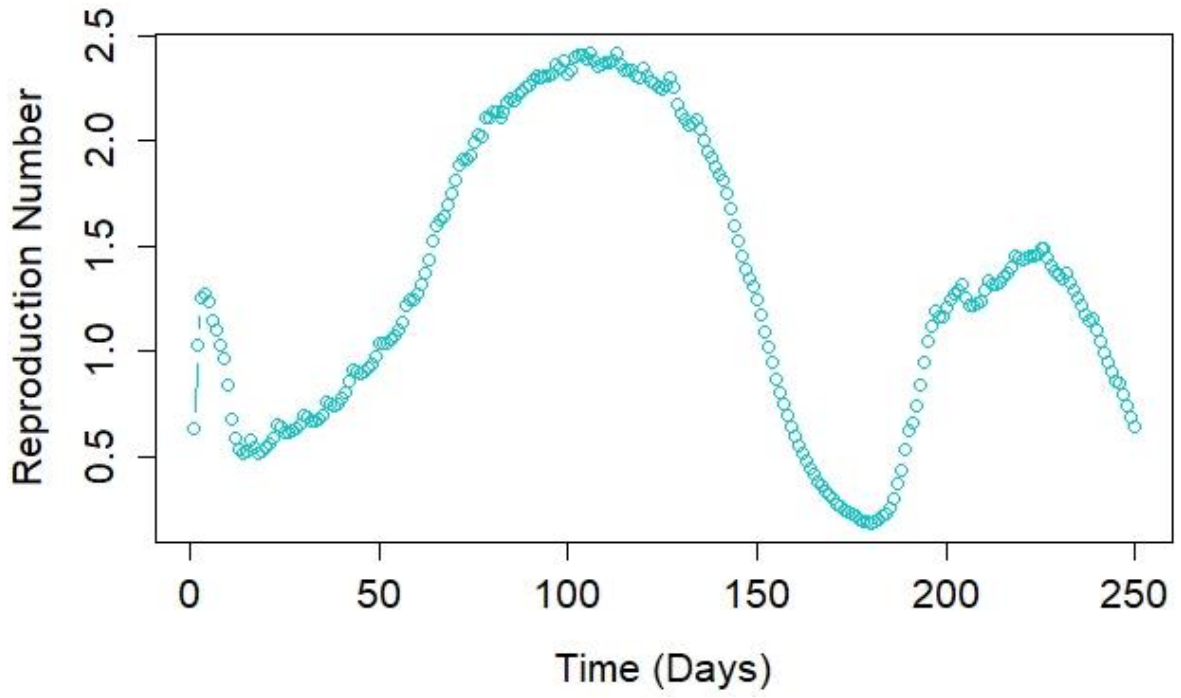

**Figure 11.** Evolution of the reproduction number during 250 days after the first reported fully vaccinated individuals

Finally, we test the predictive efficiency of the proposed model on new observations. We use the NRMSE that constitutes a reliable indicator of the predictive ability of the model, as for values less than one the proposed model provides better estimations in comparison with the historic mean of the timeseries. In figure 12, we present the corresponding estimations of the dynamic SEIHCRDV-UKF model for 15 days ahead. All NRMSE values for the 6 observable states seem highly promising, as these values are quite close to zero, emphasizing the appropriateness of our model in describing and modelling real epidemic data.

More specifically, we achieve NRMSE values of 0.26998, 0.41405, 0.35749, 0.45277, 0.39823 and 0.17426 for the number of infectious, hospital and ICU admitted, recovered, deceased and vaccinated cases, respectively. The highest NRMSE value corresponds to the recovered cases, where the dynamic SEIHCRDV – UKF model displays a slight underestimation as we move further into future states. This underestimation is influenced by the minor overestimations in the hospitalized and ICU admitted cases, as these abovementioned states are linearly associated in the SEIHCRDV model. Hence, all 6 NRMSE values are much smaller than 1, leading to the conclusion that our model can efficiently handle the prediction of future states – even for half a month ahead – especially for the number of infectious cases, hospital, ICU admissions and deaths, which are the most important variables in assessing the severity of the pandemic.

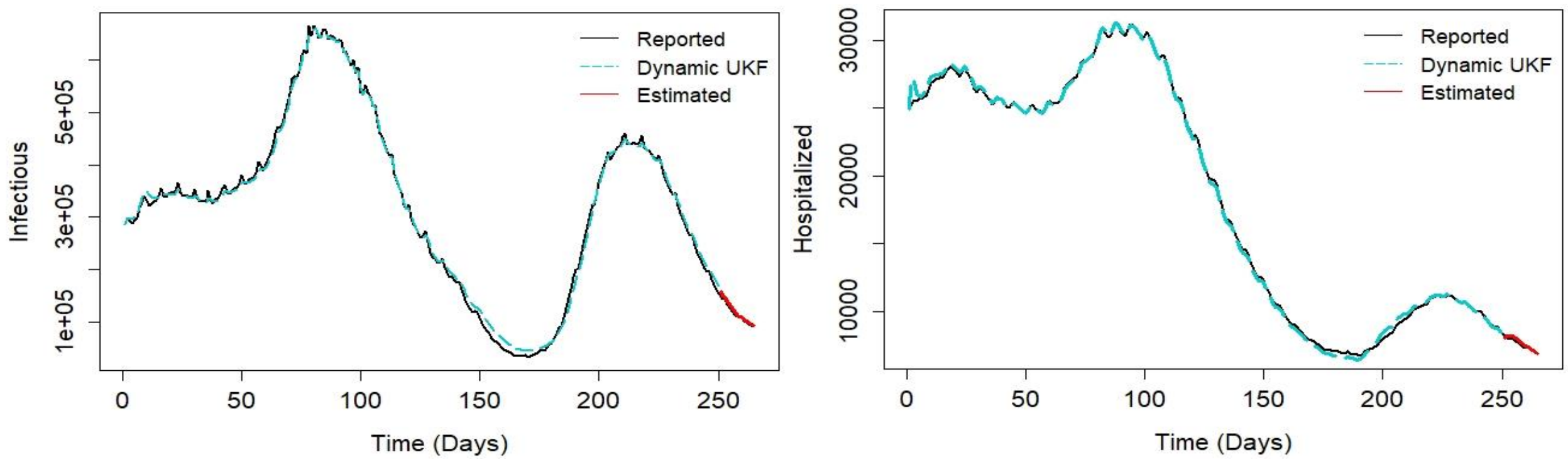

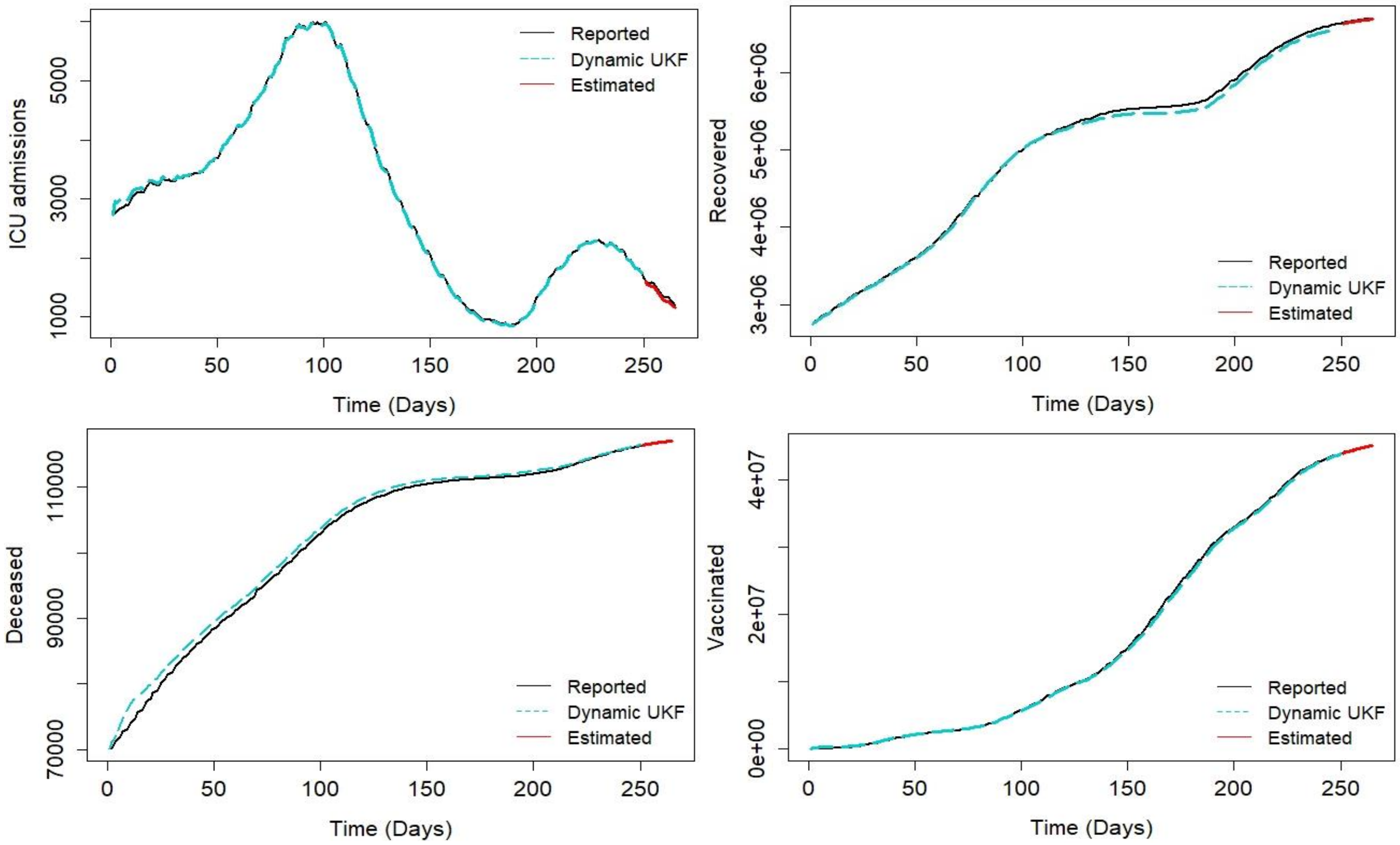

**Figure 12.** Predictive efficiency of the dynamic SEIHCRDV-UKF for half a month ahead

## 4. Discussion

Epidemics and more specifically the new SARS-CoV-2 coronavirus, which has spread rapidly around the world, cause serious problems for many national health systems. Usually, mathematical modelling of contagious diseases relies on deterministic approaches, which often use a slight extension of the standard SIR model, by adding differential equations for the exposed and deceased cases. However, this deterministic approach fails to handle the complex dynamics of COVID-19, as the parameters of the models vary significantly depending on the introduction of health restrictive measures such as lockdowns and masks, or the emergence of variants. In addition, the existence of uncertainties in the reported data is justified by the asymptomatic infections, which are difficult to detect, and the false positive-negative ratio of the PCR test. Therefore, we consider the establishment of a dynamic model that can manage these alterations.

In this paper, we introduce a novel, hybrid SEIHCRDV-UKF model with dynamic parameter estimation, aiming to address all the mentioned restrictions of the deterministic equivalent compartmental models while enhancing their predictive capacity dramatically. We extend the standard SIR model by increasing adequately the number of differential equations of the system to eight, taking into account exposed, hospital and ICU admitted, deceased and vaccinated cases. The inclusion of the hospitals and ICU admitted states in the model has an additional major advantage. In the case of COVID-19 characterized by high percentages of asymptomatic individuals, the reported daily active and recovered cases could be considered as low accuracy indices of the evolution of the pandemic. On the other hand, individuals admitted to hospital and ICUs are tested thoroughly for COVID infection, rendering the daily observations of these two states the most trustworthy data to validate the fitting-predictive ability of the proposed model.

The additional states we have included in the classical SEIR model, leading to the extended compartmental version proposed in this article, have been carefully selected aiming to encompass only states with available daily recordings. Except from the states of susceptible and exposed, for the remaining 6 states that take place in our model, there are valuable daily observations, which we efficiently deploy providing supplementary information to our stochastic epidemiological model while significantly enhancing its fitting and predictive ability.

The UKF methodology deploys the daily reported observations for a more drastic treatment of the parameter variation, leading to a stochastic approach. The inclusion of model parameters in the filtering process helps us follow the alterations in pandemics. The presented results of fitting and forecasting the spread of the disease in France – even for half a month ahead – confirm the trustworthiness of our model. We argue that this stochastic approach is necessary, as errors in the daily reported observations are known, and encapsulating all possible transitions of the pandemic would lead to a very complex model, making the process computationally expensive and the fitting-predictive performance subject to overfitting (Papageorgiou et al. 2022). For example, transitions characterized by negligible rates, such as the transitions from infection state directly to ICU admission state or from infection state directly to death state, are included in the system's additive noise. Having tested the addition of infection rates concerning exposed, hospitalized and ICU admitted cases, resulted in similar or slightly worse fitting efficiency. The parameter initialization process in non-linear methodologies should be performed with great care. Dynamic parameter estimation effectively addresses this necessity, enhancing the trustworthiness of our model.

The SEIRD model is one of the most frequently used models to describe an epidemic, containing the most common observable states. Also, as we mention in the introduction, Zhu et al. (2021) and Song et al. (2021) proposed SEIRD-EKF models with parameter estimation for the evolution of COVID in the population. As a result, we compared our model with a state-of-art model that approximates our methodology as closely as possible. The SEIRD-EKF model, based on Table 3, exhibited increased NRMSE values for the infectious, recovered and deceased cases compared to the proposed SEIHCRDV-UKF model.

The simulations presented in the results section, help us understand the severity of certain important parameters such as the re-susceptibility rate or the increased hospital and ICU admission rate that can be introduced by COVID variants, like alpha, beta or omicron. The simulation results in paragraph 3.2., for four different vaccination rates, validate the effectiveness and the benefits of vaccination against the adverse effects of COVID-19. Even a small increase in the vaccination rate would have a great benefit in controlling the coronavirus, while the effects of the vaccination are noticeable during the two infection waves of the period under consideration. This conclusion is in agreement with numerous publications that underline the importance of the vaccination campaign, as we culminate in lower mortality and hospitalization rates. For example, Watson et al. (2022) mention that the vaccination campaign resulted in a 63% reduction in expected total deaths, while Moghadas et al. (2021) state that non-intensive care unit hospitalizations and ICU hospitalizations decreased by 63.5% and 65.6%, over a period of 300 days in United States. Jabłonska et al. (2021) declare that the vaccination efficacy in protecting against deaths was 72%, with a lower reduction of the number of deaths for B.1.1.7 vs non-B.1.1.7 variants (70% and 78%, respectively). Finally, according to Dye (2022), the mortality rates in counties with low (10-39%), medium (40-69%), and high (≥70%) percentages of adults (≥18 years) who had received at least one dose of vaccine in the first half of 2021, decreased by 60%, 75%, and 81% respectively.

The theoretical analysis based on the proposed SEIHCRDV model, results into the revelation of valuable conclusions about the spread of the pandemic. Therefore, this article focuses on both the mathematical properties and the predictive efficiency of the stochastic epidemiological methodology. The derived formula for $R_0$ provides a more representative picture about the spread of pandemic, while we draw important conclusions about its evolution based on the value of $R_0$. During the 250 days, we get a reproduction number that ranges between 0.737 (1st quartile) and 1.941 (3rd quartile). The reproduction number decreases during lockdown periods and increases after the relaxation of restrictive measures. Lonergan and Chalmers (2020) estimate a $R_0$ of 2 and 0.7 for France, before and after the introduction of the first lockdown, respectively, while Glass (2020) proposes a $R_0$ of 2.02 and 0.73 during the pre- and post-lockdown period of 2020. However, these studies provide estimations prior to the time interval we investigate in our analysis. Finally, Rosero-Bixby and Miller (2022) suggest an effective reproduction number for Europe ranging between 0.8 and 1.2 during January and March 2021, which is quite close to our estimate for France over the same time period.

The presented methodology can be easily applied to many other existing or new epidemics such as ebola, infuenza, yellow fever etc., as the proposed states and transitions are representative for most of them. At the same time, this methodology is ideal in cases where both the dynamics of the system and the corresponding real-time observations cannot accurately capture the spread of an epidemic due to uncertainties, aiming to eliminate noises affecting the evolution of the states and the observations.

## 5. Conclusion

This article establishes a novel hybrid epidemiological-unscented Kalman filter model with dynamic parameter estimation that takes into account the populations of susceptible, exposed, infectious, hospitalized, ICU admitted, recovered, deceased and vaccinated cases, providing trustworthy fitting and prediction, even for half a month ahead. We note that this new consideration could be useful for examining another pandemic as well. The state-space augmentation that encapsulates 7 time-varying parameters in the updating procedure provides a much more reliable description of the spread of the pandemic in France, over a long period of 265 days. The employment of UKF, supplies the model with all available information derived from daily recordings, presenting the fluctuating dynamics of the pandemic's spread in the population. The emergence of two infection waves in this period reflects more challenging dynamics; nevertheless, our model is able to successfully deal with these challenging dynamics, due to real time parameter estimation. The proposed model outperforms not only deterministic approaches but also state-of-the-art noisy stochastic models that employ Kalman filters, such as the SEIRD-EKF, providing smaller NRMSE values.

The mathematical analysis performed according to the novel compartmental model, offers valuable results and conclusions about the nature of the pandemic. The construction of a more representative basic and effective reproductive number will also help predict and prevent outbreaks of infection, while the exploration of equilibria and their stability informs us about the nature of the disease. The simulations provided in our paper, help us monitor the severity of an increased re-susceptibility or ICU admission rate, and reveal the importance of vaccination for COVID-19. As vaccination rates increase, we end up with increasingly reduced levels of mortality and hospital admissions, while reduced re-susceptibility rates lead to shorter durations between waves of infection.

As future work we think it would be interesting to consider a hybrid epidemiological particle filter, as particle filtering constitutes another approach that deals with the uncertainty that accompanies the equations of states and observations of a phenomenon such as a pandemic. Other interesting approaches would be the implementation of a Tobit-Kalman filter (Loumponias and Tsaklidis 2022) or a Kalman Filter with non-negativity constraints (Theodosiadou and Tsaklidis 2021), due to the specificities of the modelling of epidemic outbreaks. In addition, investigating the evolution of a disease using other stochastic approaches such as discrete or continuous time Markov chains is also of great interest. Using Markov processes based on SEIHCRDV model, or a modification of it, would help us explore stochastic properties such as the expected time to hospital admission of a susceptible individual, the average duration of an infection wave, the expected number of deceased cases during a wave etc. These attempts could be the aim of forthcoming analysis.

Estimating time-dependent model parameters provides valuable information about the future of the pandemic. The results presented are encouraging, given that as we move deeper into the vaccination period, mortality and hospital admission rates show a steady downward trend even if the transmissibility of the disease does not decrease after 2.5 years of pandemic. COVID's $R_0$ is much higher than 1, and data shows that its value can be diminished only by decisive interferences such as lockdowns. Given this situation – at least for the foreseeable future – national health systems should emphasize the deterioration of the severely infected and death rates, as the limitation of the contagiousness of the disease seems quite difficult so far, revealing once again the importance of full vaccination in the population. Finally, the encouraging indicators encountered so far should not lead to relaxation of the awareness of health systems, because there is still a non-negligible death rate, post-COVID conditions are making their presence felt, while new mutations could dramatically change the present landscape.

## Declarations

**Funding:** Not applicable
**Conflicts of interest/Competing interests:** The authors declare that there is no conflict of interests.